\RequirePackage{fix-cm}
\documentclass[smallextended]{svjour3}       % onecolumn (second format)
\smartqed  % flush right qed marks, e.g. at end of proof
\usepackage{graphicx}
\usepackage{array}
\usepackage{amsmath}
\usepackage{amsfonts}
\usepackage{subfig}
\usepackage{float}
\usepackage{cite}
\usepackage[colorlinks=true, allcolors=blue]{hyperref}
\usepackage{xcolor}

\DeclareMathOperator{\sech}{sech}
\begin{document}

\title{Bright soliton interactions in the variable coefficient Fokas-Lenells equation, Conservation laws, Modulation instability and Soliton tunneling%\thanks{Grants or other notes
}

\titlerunning{Variable coefficient Fokas-Lenells equation}        % if too long for running head

\author{Sagardeep Talukdar \and R. Ramakrishnan \and Sudipta Nandy       \and  M. Lakshmanan
}


\institute{Sagardeep Talukdar \at 
	Department of Physics, Cotton University, Guwahati 781001,\\
	Assam, India, \\
	\email{talukdarsagardeep@gmail.com}          %  \\
           \and
          R. Ramakrishnan (Corresponding Author) \at
          Department of Physics, Indian Institute of Technology Kharagpur, Kharagpur,\\ 
          West Bengal - 721302, India,\\
          \email{ramakrishnan.cnld@gmail.com}  
          \and
          Sudipta Nandy  \at
          Department of Physics, Cotton University, Guwahati 781001,\\
          Assam, India, \\
          \email{sudiptanandy@gmail.com} 
           \and
          M. Lakshmanan  \at
          Department of Nonlinear Dynamics, Bharathidasan University,
          Tiruchirappalli-620 024,\\
          Tamil Nadu, India\\
          \email{lakshman.cnld@gmail.com}   
            }

\date{Received: date / Accepted: date}

\maketitle

\begin{abstract}
We present here a study of the bright soliton dynamics in an inhomogeneous fibre by means of variable coefficient Fokas-Lenells equation with time varying dispersion, nonlinearity and gain/loss parameter. At first we propose our system that governs the propagation of ultrashort pulses in inhomogeneous fibre. Secondly, under a suitable gauge transformation we transform the system into a simplified form of variable coefficient Fokas-Lenells equation. The Lax integrability and conservation laws are exhibited. We also study the stability of the generalized plane wave against small amplitude perturbation. Thereafter, by using a nonstandard Hirota bilinearization method with the help of suitable auxiliary function, we obtain the bright one soliton, two soliton and provide a scheme for obtaining $N$-bright soliton solutions. The elastic collision dynamics of the two solitons is studied using asymptotic analysis. We also investigate the soliton acceleration/retardation under suitable choice of dispersion and nonlinearity coefficients. Finally, the dramatic effect of nonlinear tunneling of the bright one and two soliton is also studied under some Gaussian dispersion or nonlinearity. \\    
%\subclass{MSC code1 \and MSC code2 \and more}
\end{abstract}

\section{Introduction}
\label{intro}
\indent An optical soliton is a nonlinear pulse that travels through an optical fibre without distortion, making it an ideal candidate for transporting data through optical fibres in digital and quantum communication systems. In nonlinear optical fibres, group velocity dispersion (GVD) and nonlinearity are very crucial to the propagation of these solitons. This unique property of solitons arise due to a balance between GVD, nonlinearity, and the gain or loss within the optical fibre. The history of these optical solitons dates back to the year 1973, when Hasegawa and Tappert \cite{Hasegawa} theoretically predicted their existence. Seven years later, in 1980, it was successfully realised experimentally\cite{mollenauer1980experimental}. Since then, research on optical solitons has consistently been a topic of significant interest. Besides wave propagation, the notion of soliton also plays a significant role in cavitation dynamics, where the collapse of soliton-like bubbles can generate highly localized energy release, enabling phenomena such as shock emission and micro-bot formation \cite{wang2025launching}. Moreover very recently increasing attention has also been devoted to multi-pole solitons arising out of fractional nonlinear evolution equations. Solitons arising from these fractional models have demonstrated potential applications across various fields\cite{xu2025multipole}. It is this rich and diverse dynamical behavior of solitons across various physical disciplines that continues to make them an important and attractive subject of study, drawing sustained interest from researchers in both theoretical and applied sciences.\\
 \indent Over the past few decades, the study of integrable nonlinear dynamical systems governing the propagation of these optical pulses has been done extensively. Among the various integrable systems in nonlinear optics, the nonlinear Schr\"odinger equation(NLSE) \cite{Hasegawa,serkin2000novel}, higher order NLSE(HNLSE) \cite{SSE} and derivative NLSE(DNLSE)\cite{Kaup} are the fundamental models. Solitons arising from these systems find broad applications across various fields, including nonlinear optics \cite{kivshar2003optical,agrawal2000nonlinear,hasegawa1995solitons}, plasma physics \cite{zabusky1965interaction} and hydrodynamics \cite{kuznetsov1986soliton}, photonic Moir\'{e} lattice\cite{mou2025optical} etc.\\   
\indent However, real-world systems are often inhomogeneous due to medium irregularities in optical fibres. It is seen that in many cases the soliton parameters, namely pulse width, chirp, and position, may vary drastically from their initial values with the increase in the dispersion and nonlinear effects. So, to address these physical variations, many authors have studied the dynamics of nonautonomous solitons, i.e., solitons arising from evolution equations with spatially or temporally varying coefficients. Such a concept was first introduced by Serkin and Haswgawa \cite{serkin2000novel,serkin2002exactly} which led to the beginning of dispersion managed (DM) solitons. After this, the concept of soliton dispersion management became prominent in pulse propagation through optical fibres. This enables one to mitigate the challenges faced in soliton transmission by strategically designing fibre dispersion by connecting fibres with different dispersion values. In the context of NLSE, the DM soliton is well studied under different settings\cite{serkin2000novel, serkin2002exactly, serkin2007nonautonomous, serkin2010solitary, kumar2021dispersion}. %Moreover, in the context of Bose-Einstein condensate, the nonautonomous bright soliton is obtained in \cite{triki2020nonautonomous}.
But, when it comes to high intensity light beams that produce ultrashort (femtosecond) pulses one must consider higher order generalization of NLSE that describes additional physical effects, namely spatio-temporal dispersion, nonlinear dispersion, etc. One such generalization of NLSE is the Fokas-Lenells equation (FLE) that was first proposed in refs. \cite{Fokas,Lenells}. The FLE in dimensionless form is
\begin{align}\label{FLE}
	iQ_T  + \gamma_1 Q_{XX} + \gamma_3 |Q|^2 Q + \gamma_2 ( i \gamma_3  |Q|^2Q_X-Q_{XT}) = 0, 
\end{align}   
where $Q$ is  the  envelope function of an optical field. The second and third terms of Eq. (\ref{FLE}) are the group velocity dispersion term and cubic nonlinearity term, respectively. The second last term in Eq. (\ref{FLE}) is the intensity dependent contribution to the group velocity which in turn leads to the phenomena of self-steepening\cite{agrawal2000nonlinear} and the last term is the spatio-temporal dispersion term. The FLE, introduced by Fokas \cite{Fokas} and subsequently developed by Lenells \cite{Lenells}, represents a significant integrable extension of the NLSE. In \cite{Fokas}, the model was established within the bi-Hamiltonian framework, while in \cite{Lenells} its physical foundation was clarified. The FLE belongs to the hierarchy of derivative NLSE-type systems and is associated with the first negative hierarchy. An important reduction occurs when $\gamma_2 = 0$, in which case the equation simplifies to the well known classical NLSE.\\ 
\indent  The integrable structure of the FLE has been extensively explored. Lenells and Fokas \cite{FL} derived its conservation laws and Lax pair using the underlying bi-Hamiltonian structure, and further solved the initial value problem through the inverse scattering transform method. Following these pioneering works, a wide variety of localized wave solutions including solitons, breathers, and rogue waves have been obtained via various analytical techniques such as the inverse scattering technique\cite{lashkin2021perturbation}, Darboux transformation \cite{he2011self,chen2014peregrine}, the dressing method \cite{lenells2010dressing} and Hirota's bilinear approach \cite{Matsuno1,Matsuno2,talukdar2023multi,dutta2023fokas,lu2012novel,liu2022fokas}.\\
\indent However, the study on the propagation dynamics of ultrashort pulses modeled by FLE in the soliton management schemes i.e. with spatially/temporally varying dispersion and nonlinearity is relatively sparse. Only few works have been reported in the literature that discusses the nonautonomous soliton dynamics of FLE\cite{kundu2010two,lu2013nonautonomous,silem2023nonautonomous,wang2017higher}. However, to the best of our knowledge, in the dispersion and nonlinearity regime with gain/loss parameter the FLE is not yet studied. Such a soliton management scheme finds a prominent place in the study of solitons as it enriches the propagation dynamics of the soliton, allowing one to manipulate it, and also allows the study of additional aspects namely soliton tunneling\cite{serkin2001nonlinear,serkin2013geiger}, etc. Thus, in this manuscript we introduce the FLE with modified dispersion, nonlinearity and gain/loss coefficients as follows,
\begin{align}\label{FLE2} \nonumber
	iQ_T  + \gamma_1 D(T) Q_{XX} + \gamma_3R(T) |Q|^2 Q + \gamma_2 ( i \gamma_3 R(T) |Q|^2Q_X-Q_{XT}) \\ = \frac{\gamma_1}{2\gamma_2^3}\Gamma(T)(\gamma_2Q_X-iQ). 
\end{align} 
Here, the coefficient $D(T)$ is the dispersion coefficient, $R(T)$ is the nonlinearity coefficient and $\Gamma(T)$ is the gain/loss parameter. The coefficients $D(T)$ and $R(T)$ allow us to modulate the dispersion and nonlinearity with time. Based on the choice of dispersion and nonlinearity coefficients the parameter $\Gamma(T)$ takes different forms and is responsible for the gain/loss in the amplitude of the soliton. The forms of $\Gamma(T)$ are discussed later.\\
We recall the standard gauge transformation used in \cite{dutta2025soliton} as follows,
\begin{align}
	Q = \sqrt{\frac{m}{|\gamma_{3}|}} \,
	n e^{i\left(nX + 2mn \int D(T)\, dT \right)} q,
\end{align}
where $\gamma_{3}=\pm 1$, $m=\frac{\gamma_{1}}{\gamma_{2}}$ and $n=\frac{1}{\gamma_{2}}$. Now the above transformation converts Eq. (\ref{FLE2}) into the following form,
\begin{align}
	- Q_{XT} + m D(T) Q_{XX} - m n^{2} \big( D(T)Q - i R(T) |Q|^{2} Q_{X} \big)
	= \frac{mn^2}{2} \Gamma(T) Q_{X}.
\end{align}
Following the transformation of the variables as, 
\begin{align}
	x = 2 \left( X + m \int D(T)\, dT \right), 
	\qquad 
	t = -\frac{m n^{2}}{2}\, T,
\end{align}
and time varying coefficients as,
\begin{align}
	D(T) \to D(t), 
	\qquad 
	R(T) \to R(t), 
	\qquad 
	\Gamma(T) \to \Gamma(t),
\end{align}
\begin{align}
	D(t) = -\frac{m n^{2}}{2}\, D(T), 
	\qquad 
	R(t) = -\frac{m n^{2}}{2}\, R(T), 
	\qquad 
	\Gamma(t) = -\frac{m n^{2}}{2}\, \Gamma(T),
\end{align}
we obtain the following variable coefficient FLE (vcFLE) model,
\begin{align}\label{model}
	&q_{xt} - D(t) q +2i  R(t) |q|^2 q_x   =  \Gamma(t)q_x .
\end{align}
Here, $q$ is the newly introduced complex field. The term $\Gamma(t)$ takes the form as 
\begin{align}
	\Gamma(t)=\frac{W[R(t),D(t)]}{2\ R(t)\ D(t)},
\end{align} 
where the Wronskian is $W[R(t),D(t)]=R(t)D'(t)-D(t)R'(t)$. The $'$ denotes derivative with respect to time. This very form of $\Gamma(t)$ should be preserved so that Eq. (\ref{model}) is integrable which will be discussed in the next section. The gain/loss parameter is dependent on the two coefficients $D(t)$ and $R(t)$ by the following ways i.e. in the first case if the dispersion coefficient \( D(t) \) and the nonlinear coefficient \( R(t) \) are linearly dependent, the gain/loss term \( \Gamma(t) \) vanishes, indicating that the soliton experiences no net amplification/loss. For example, the choice \( D(t) = e^{-t^2} \) and \( R(t) = c_1 e^{-t^2} \), where $c_{1}$ is a constant, leads to \( \Gamma(t) = 0 \). In the second case, when \( D(t) \) is proportional to \( R(t) \) through some function, \( \Gamma(t) \) becomes constant, resulting in a steady gain or loss; for instance, if \( D(t) = c_1 e^{-t} \), \( R(t) = 1 \), then \( \Gamma(t) = \frac{1}{2} \). Thirdly, a more general situation arises when \( D(t) \) and \( R(t) \) are linearly independent, in which case \( \Gamma(t) \) is time-dependent, leading to a nonuniform gain or loss. For example, with \( D(t) = b e^{-c t^2} \) and \( R(t) = 1 \), one obtains \( \Gamma(t) = c t \), which varies explicitly with time. One can see that, when we consider $D(t)=R(t)=1$, then the gain/loss parameter $\Gamma(t)$ becomes zero which is the well known constant coefficient FLE. Notice that the three cases make the study of the nonautonomous dynamics of soliton solution of vcFLE given by Eq. (\ref{model}) more interesting and rich.  \\ 
\indent Therefore, in this manuscript, our purpose is to study the vcFLE obtained in Eq. (\ref{model}) by understanding the underlying nonautonomous soliton dynamics. Thus in Sec. 2 we present the Lax pair and derive the conserved quantities. Then in Sec. 3 we perform the linear stability analysis of the generalized plane wave solution admitted by the system. Subsequently, we shall obtain the bright one and two soliton solutions and provide the scheme for obtaining the $N$-soliton solutions in Sec. 4. Next, in Sec. 5 we shall discuss the interaction dynamics and the phenomenon of soliton tunneling under some physical choice of dispersion and nonlinearity. Finally, in Sec. 6, we provide a brief conclusion.\\  
\section{Associated spectral problem and conservation laws}
\label{LAX}
\indent The Lax pair $(U,V)$ associated with Eq. (\ref{model}) is given by,
\begin{align}
	\label{LaxFL}
	\partial_x { \Psi} = U { \Psi}, \\
	\partial_t { \Psi} = V { \Psi}.
\end{align}
Here, $\bf{\Psi}$ is a two component vector field, expressed as,
\begin{align}
	\label{Psi}
	{ \Psi}= ({\Psi_1} \ {\Psi_2})^T.
\end{align} 
Also, $U$ and $V$ are given by,
\begin{align}
	\label{Lax}
	U=&\left(
	\begin{array}{cc}
		\frac{-i}{2}  \Lambda ^2 & \Lambda  q_x \sqrt{\frac{R(t)}{D(t)}} \\
		-\Lambda  q^{*}_{x} \sqrt{\frac{R(t)}{D(t)}} & \frac{i}{2}\Lambda ^2 \\
	\end{array}
	\right) ,     \quad
	V=&\left(
	\begin{array}{cc}
		A & B \\
		C &-A \\
	\end{array}
	\right),
\end{align} 
where $A=\frac{i D(t)}{2 \Lambda ^2}-i |q|^2 R(t)$, $B=-\frac{i q D(t) \sqrt{\frac{R(t)}{D(t)}}}{\Lambda }$, $C=	-\frac{i q^{*} D(t) \sqrt{\frac{R(t)}{D(t)}}}{\Lambda }$. $\Lambda$ is the spectral parameter and $q$, $q^{*}$ are complex valued and are functions of $x$, $t$. The compatibility equation of $(U,V)$, namely, the zero curvature equation $U_t -V_x + [U,V] = 0$, provides Eq. (\ref{model}) which contains a new additional term proportional to $\Gamma(t)$.\\ 
Now, writing ($U,V$) given in Eq. (\ref{Lax}) in equation form, we get
\begin{align}
	\label{Lax2}
	U =& \frac{-i\Lambda^2 }{2} \ \sigma   +\sqrt{\frac{R(t)}{D(t)}} \Lambda \  u_x, \\
	V =& \frac{iD(t) }{2  } \Lambda^2 \ \sigma -  \frac{i}{\Lambda}\ \sqrt{D(t)R(t)} \sigma u + iR(t)\sigma u^2,
\end{align} 
where $\sigma=\left(
	\begin{array}{cc}
		1 & 0 \\
		0 &-1 \\
	\end{array}
	\right)$, and   
	$u=\left(
	\begin{array}{cc}
		0 & q \\
		-q^{*} & 0 \\
	\end{array}
	\right)$. 
The Lax Eq. (\ref{Lax2}) in component form is,
\begin{align}
	\label{component1}
	\Psi_{1x} & = \frac{-i \Lambda^2}{2} \Psi_1  + \sqrt{\frac{R(t)}{D(t)}} \Lambda q_x \Psi_2,\\
	\label{component2}
	\Psi_{2x} & = -\sqrt{\frac{R(t)}{D(t)}}\Lambda \ q_x^* \ \Psi_1 + \frac{i \Lambda^2}{2} \Psi_2.  
\end{align}
Now following a similar procedure as in \cite{Ghosh1999} we write 
\begin{align}
	\gamma = \frac{\Psi_1}{\Psi_2}.
\end{align}
Then from Eqs. (\ref{component1}) and (\ref{component2}),  we obtain a first order  nonlinear differential equation,
\begin{align}
	\label{Riccati2}
	\gamma_x = -i \Lambda^2 \gamma + \sqrt{\frac{R(t)}{D(t)}}\Lambda q_x +\sqrt{\frac{R(t)}{D(t)}} \Lambda q_x^* \gamma^2,
\end{align}
which is known as Riccati equation. The solution of the Riccati Eq. (\ref{Riccati2}) is related to the conserved quantities in the following way, 
\begin{align}
	\label{Conserve-1}
	\ln(a_{22}(\Lambda))  =  \ln (e^{- i \frac{\Lambda^2}{2} x} \Psi_2 )|_{x \rightarrow \pm \infty} 
	= -\Lambda \int_{-\infty}^{\infty} q_x^* \gamma \ dx= \sum_{n=0}^{\pm \infty} i^{n} H_n \Lambda^{-n+1}.   
\end{align}
The quantity  $ a_{22}$ in Eq. (\ref{Conserve-1}) is the scattering parameter and is time independent. In the summation, the negative powers of $\Lambda $ give  positive hierarchy $H_n$,  whereas the positive powers of $\Lambda $ give negative hierarchy $ H_{-n}$. However, here in this paper we only show the first three conserved quantities in the positive hierarchy \cite{mlbook}. Consider that the Eq. (\ref{Riccati2}) has a series solution, 
\begin{align}
	\gamma = \sum_{n=0}^{\infty} a_{\pm n} \Lambda^{\mp n}.
\end{align}
Now, by substituting $\gamma$ in Eq. (\ref{Riccati2}) we obtain the coefficients $ a_n $ from the coefficients of $\Lambda$ as, 
\begin{align}
	a_0 & =0,\quad a_1= - i \sqrt{\frac{R(t)}{D(t)}} q_x, \\
	a_2 &=0, \quad a_3 =\sqrt{\frac{R(t)}{D(t)}}\bigg(q_{xx} + i \frac{R(t)}{D(t)} |q_x|^2 q_x\bigg),\\
	a_4 &=0, \quad a_5=\sqrt{\frac{R(t)}{D(t)}}\bigg[i (q_{xxx}-2\ \frac{R(t)^2}{D(t)^2}|q_x|^4 q_x ) 
	-4\ \frac{R(t)}{D(t)}|q_x|^2q_{xx} -\frac{R(t)}{D(t)} q_x^2 q_{xx}^*\bigg].
\end{align}
Thus, the conserved quantities obtained are,
\begin{align}
	H_1 &= \sqrt{\frac{R(t)}{D(t)}}\int_{-\infty}^{\infty} |q_x|^2 dx, \\
	H_3 &= \sqrt{\frac{R(t)}{D(t)}}\int_{-\infty}^{\infty} \bigg( \frac{R(t)}{D(t)}|q_x|^4   -i q_x^* q_{xx} \bigg) dx,  \\ \nonumber
	H_5 &=\sqrt{\frac{R(t)}{D(t)}} \int_{-\infty}^{\infty} \bigg(- q_x^* q_{xxx} + 2\ \frac{R(t)^2}{D(t)^2}|q_x|^6 -4i \frac{R(t)}{D(t)}|q_x|^2 q_x^* q_{xx}\\
	&-i\frac{R(t)}{D(t)}|q_x|^2 q_x q_{xx}^*  \bigg) dx.
\end{align}
%%%%%%%%%%%%%%%%%%%%%%%%%%%%%%%%%%%%%%%%%%%%%%%%%%%%%%%%%%%%%%%%%%%%%%%%%%%%%%%%%%%%%%%%%%%%%%%%%%%%%%%%%%%%%%%%%%%%%%%%%%%%%%%%%%%%%%%%%%%%%%%%%%%%%%%%%%%%%%%%%%%%%%%%%%%%%%%%%%%%%%%%%%%%%%%%%%%%%%%
\section{Modulation instability analysis}
\indent In nonlinear systems, dispersion and nonlinearity influence the system in distinct ways, yet both can trigger instabilities. In fibre optical models governed by nonlinear equations, even a stable solution may become unstable due to the combined effects of these two terms, and such a phenomenon is known as modulation instability. This process occurs in nonlinear dispersive media when a continuous plane wave undergoes self-modulation in amplitude and frequency, causing small perturbations to grow exponentially \cite{agrawal2000nonlinear}. Hence in this section, we investigate the linear stability analysis of the generalized plane wave for the vcFLE given by Eq. (\ref{model}) against small perturbation. We remark that the stability analysis of the constant coefficient FLE on the plane wave background is already discussed in \cite{Matsuno2}. However, we intend to investigate whether additional dispersion and nonlinear coefficients in Eq. (\ref{model}) affect the stability of the generalized plane wave.\\
\indent The generalized plane wave solution of Eq. (\ref{model}) is given as,
\begin{align}
	q_{pw}=\sqrt{\frac{D(t)}{R(t)}} A\ e^{i(K_1 x+\omega_1(t))},
\end{align}  
where $\sqrt{\frac{D(t)}{R(t)}}A$ is the generalized plane wave amplitude, $K_1$ is the wave number, $\omega_1(t)/t$ is the generalized frequency. The dispersion relation $\omega_1(t)=-\int \frac{2 K_1  \text{A}^2 +1}{K_1} \,D(t) dt$ ensures $q_{pw}$ is the solution of Eq. (\ref{model}).\\
\indent Following the standard procedure of linear stability analysis we seek the perturbed solution of the form,
\begin{align}\label{pw}
	q_{pw}=\sqrt{\frac{D(t)}{R(t)}}(A+\Delta A)\ e^{i(K_1 x+\omega_1(t))}.
\end{align}	 
Here, $\Delta A=\Delta A(x,t)$ is a small perturbation added to the amplitude of the generalized plane wave with
\begin{align}
	\Delta A(x,t)=m e^{i(K_2x+\omega_2(t))}+ne^{-i(K_2x+\omega_2(t))},
\end{align}
where $K_2$ and $\omega_2(t)/t$ are the perturbed wave number and generalized frequency respectively. Also, $m$, $n$ are two small real parameters. Substituting the perturbed plane wave given by Eq. (\ref{pw}) in Eq. (\ref{model}) we find a system of homogeneous equations for $m$ and $n$ as follows,
\begin{align}
	\left(
	\begin{array}{cc}
		\frac{\left(\text{$K_2 $}-2 A^2 \text{$K_1$}^2\right) \text{D}(t)-\text{$K_1$} (\text{$K_1$}+\text{$K_2$}) \text{$\omega_2$}'(t)}{\text{$K_1$}} & -2 A^2 \text{$K_1$} \text{D}(t) \\
		-2 A^2 \text{$K_1$} \text{D}(t) & \frac{\text{$K_1$} (\text{$K_1$}-\text{$K_2$}) \text{$\omega_2$}'(t)-\left(2 A^2 \text{$K_1$}^2+\text{$K_2$}\right) \text{D}(t)}{\text{$K_1$}} \\
	\end{array}
	\right)
	\left(
	\begin{array}{c}
		\text{$m$} \\
		\text{$n$} \\
	\end{array}
	\right)=0.
\end{align}
To obtain the solutions of the homogeneous equations we consider,
\begin{align}
	\left|
	\begin{array}{cc}
		\dfrac{\left(K_2 - 2 A^2 K_1^2\right) D(t) - K_1 \left(K_1 + K_2\right) \omega_2'(t)}{K_1} & -2 A^2 K_1 D(t) \\[1em]
		-2 A^2 K_1 D(t) & \dfrac{K_1 \left(K_1 - K_2\right) \omega_2'(t) - \left(2 A^2 K_1^2 + K_2\right) D(t)}{K_1}
	\end{array}
	\right|=0.
\end{align} 
From the above determinant we obtain the new dispersion relation as,
\begin{align}
	\text{$\omega_2$}(t)=\frac{ -\left(2 A^2 \text{$K_1$} \text{$K_2$}+\text{$K_2$}\right)\pm \frac{K_2}{K_1}\sqrt{ \left(4 A^2 \text{$K_1$}^3( A^2 \text{$K_1$}+1)+\text{$K_2$}^2\right)}}{\text{$K_2$}^2-\text{$K_1$}^2}\int D(t)dt.
\end{align}
From the above expression, it implies that $\omega_2(t)$ is real only when,
\begin{align} \label{condition}
	(4 A^2 \text{$K_1$}^3( A^2 \text{$K_1$}+1)+\text{$K_2$}^2)>0.
\end{align} 
The generalized plane wave is naturally stable as long as the above condition holds. Additionally, it is also evident that the dispersion term $D(t)$ does not influence the stability of the generalized plane wave. 
%%%%%%%%%%%%%%%%%%%%%%%%%%%%%%%%%%%%%%%%%%%%%%%%%%%%%%%%%%%%%%%%%%%%%%%%%%%%%%%%%%%%%%%%%%%%%%%%%%%%%%%%%%%%%%%%%%%%%%%%%%%%%%%%%%%%%%%%%%%%%%%%%%%%%%%%%%%%%%%%%%%%%%%%%%%%%%%%%%%%%%%%%%%%%%%%    
\section{Nonstandard Hirota Bilinearization method}
\indent In order to understand the underlying dynamics of the vcFLE system one has to solve the system given by Eq.(\ref{model}). In this regard, we apply the nonstandard Hirota's bilinearization method \cite{Hirota_2004} and rewrite  Eq. (\ref{model}) in the bilinear form. Then we obtain the bright one and two soliton solutions and provide the scheme for obtaining the $N$-bright soliton solutions. To bilinearize Eq. (\ref{model}) we consider the bilinearizing transformation, 
\begin{align}
	\label{bilin0}
	q(x,t) =\sqrt{\frac{D(t)}{R(t)}}\frac{G}{F}, \quad q^*(x,t)=\sqrt{\frac{D(t)}{R(t)}}\frac{G^*}{F^*},
\end{align}
where $G(x,t)$, $F(x,t)$ are complex functions and $G^*(x,t)$, $F^*(x,t)$ are their complex conjugates. Inserting Eq. (\ref{bilin0}) in  Eq. (\ref{model}) we obtain the following nonlinear equation(as well as its complex conjugate version) in terms of the newly introduced functions,
\begin{align}
	\label{Bilin1}
	& \frac{1}{F^2} \left(D_x D_t - D(t)\right)\, G \cdot F
	- \frac{G}{F^3}\, D_x D_t \left(F \cdot F\right)- \frac{2 D(t)\, G\, G^*}{F^3\, F^*}\, D_x \left(G \cdot F\right) \nonumber \\
	& + \frac{s\, G\, G^*}{F^3} 
	- \frac{s\, G\, G^*\, F^*}{F^3\, F^*} = 0.
\end{align}
Here $D_x$ and $D_t $ are the standard Hirota operators \cite{Hirota_2004}, which are defined as,
\begin{align}
	D_x^m D_t^n G.F= 
	(\frac{\partial}{\partial x}  - \frac{\partial}{\partial x^\prime})^m
	(\frac{\partial}{\partial t}  - \frac{\partial}{\partial t^\prime})^n
	G.F\Bigg|_{ (x=x^\prime)(t= t^\prime)},
\end{align} 
where $m$ and $n$ are positive integers. Notice that, the last two terms in Eq. (\ref{Bilin1}) contain an auxiliary function $s(x,t)$, which is introduced so that Eq. (\ref{Bilin1}) can be cast into the following bilinear equations,
\begin{align}
	\label{BR1}
	&\Big[D_x D_t -D(t)\Big]G.F=0,\\
	\label{BR2}
	&D_x D_t F.F  = s G^*,\\
	\label{BR3}
	&2 i D(t) D_x G.F   = s F^*,
\end{align}
where $D(t)$ is the dispersion coefficient as discussed earlier. A key difference lies in our approach of nonstandard Hirota bilinearization compared to that of the standard bilinearization carried out by Matsuno in \cite{Matsuno1}. Here, we have introduced an auxiliary function $s(x,t)$ to convert the trilinear equation given by Eq. (\ref{Bilin1}) into a set of bilinear equations given in Eqs. (\ref{BR1})-(\ref{BR3}). The use of the auxiliary function $s$ has simplified the conversion from the trilinear to bilinear equations giving the desired result. On the other hand, in \cite{Matsuno1}, a specific identity is employed to transform the trilinear equation into the bilinear forms. 

To solve the above given bilinear equations we expand $G$, $F$, $s$, $G^*$, and $F^*$ in terms of the following power series with an expansion parameter $\epsilon$:
\begin{align}
	\label{GF1}
	G&= \epsilon G_1 + \epsilon^3 G_3 +.....,\quad
	&&G^*= \epsilon G_1^* + \epsilon^3 G_3^* +....., \\
	\label{GF2}
	F&= 1 + \epsilon^2 F_2 + \epsilon^4 F_4 +.....,\quad
	&&F^*= 1 + \epsilon^2 F_2^* + \epsilon^4 F_4^* +....., \\
	\label{GF3}
	s&= \epsilon s_{1} + \epsilon^3 s_{3} + .....  .
\end{align}
Substituting Eqs. (\ref{GF1}-\ref{GF3}) into Eqs. (\ref{BR1}-\ref{BR3}) and collecting various powers of $\epsilon$, we obtain the following set of linear partial differential equations(PDEs).\\
From the 1st bilinear equation, Eq. (\ref{BR1}),
\begin{align}
	\mathcal{O}(\epsilon):&G_{1xt}-D(t)G_1=0,\\
	\mathcal{O}(\epsilon^3):&G_{1xt}F_2-G_{1x}F_{2t}-G_{1t}F_{2x}+G_1F_{2xt}-D(t)G_1F_2=0.
\end{align}
From the 2nd bilinear equation, Eq. (\ref{BR2}),
\begin{align}
	\mathcal{O}(\epsilon^2):&2F_{2xt}-s_1G_1^*=0,\\
	\mathcal{O}(\epsilon^4):&F_{2xt}F_2-F_{2x}F_{2t}=0.
\end{align}
From the 3rd bilinear equation, Eq. (\ref{BR3}),
\begin{align}
	\mathcal{O}(\epsilon):&2iD(t)G_{1x}-s_1=0,\\
	\mathcal{O}(\epsilon^3):&2iD(t)[G_{1x}F_2-G_{1}F_{2x}]-s_1F_2^*=0.
\end{align}
Systematically solving these linear set of PDEs with the appropriate seed solutions one can construct multi-soliton solutions.
%%%%%%%%%%%%%%%%%%%%%%%%%%%%%%%%%%%%%%%%%%%%%%%%%%%%%%%%%%%%%%%%%%%%%%%%%%%%%%%%%%%%%%%%%%%%%%%%%%%%%%%%%%%%%%%%%%%%%%%%%%%%%%%%%%%%%%%%%%%%%%%%%%%%%%%%%%%%%%%%%%%%%%%%%%%%%%%%%%%%%%
\subsection{Bright one soliton solution}
To construct the bright one soliton solution of the vcFLE system we consider the seed solution as $G_1= \alpha_1 e^{\eta_1}$ where $\eta_1=p_1 x+\Omega_1(t)$. Then, successively solving the above given system of linear PDEs, we obtained the unknown function $F_{2}$ with appropriate form of $s=c_{1} e^{\eta_{1}}$. Hence, for the bright one soliton the considered power series of $G$ and $s$ get truncated at the order $\epsilon$ whereas the series of $F$ gets truncated at the order $\epsilon^{2}$. The resultant bright one soliton solution(1SS) of the vcFLE can be written as,
\begin{align}
	\label{sol2}
	q(x,t)=\sqrt{\frac{D(t)}{R(t)}} \frac{\alpha_1\ e^{\eta_1}}{1 + \beta_1 e^{\eta_1+\eta_1^*}},
\end{align}
where 
\begin{align}
	\Omega_1(t)= \int& \frac{D(t) dt}{p_1}, \quad  \Omega_1^*(t)= \int \frac{D(t)dt}{p_1^*}, \\
	&\beta_1 = i \frac{|\alpha_1|^2 |p_1|^2 p_1}{ (p_1+p_1^*)^2}.
\end{align}
Here the parameter $\alpha_{1}$ and the wave number $p_{1}$ are arbitrary complex constants. These two complex parameters along with the functions $D(t)$ and $R(t)$ determine the properties of the bright one soliton solution of the vcFLE. Now we can parameterize the complex wave number $p_{1}=a_{1}+ib_{1}$. In terms of the new parameters, one bright soliton solution (\ref{sol2}) can be rewritten in the form
\begin{align}
q(x,t)=\sqrt{\frac{D(t)}{R(t)}} \frac{4 a_1^2 \alpha_1 e^{(\theta_1+i\chi_1)}}{4 a_1^2 + |\alpha_1|^2 (b_1 + i a_1)(a_1 + i b_1)^2 e^{2\theta_1}},
\end{align}
As a consequence the intensity of one soliton can be written as,
\begin{align}
\label{q1}
&|q(x,t)|^2=\frac{D(t)}{R(t)}\frac{2 a^2_1}{(a^2_1+b^2_1)^{3/2}}\times \frac{1}{cosh[2(\theta_1+ln\ \delta_1)]-\frac{b_1}{\sqrt{a^2_1+b^2_1}}},
\end{align}
where $\theta_1=a_1 x+\frac{a_1 \int D(t)dt}{a^2_1+b^2_1}$,  $\chi_1=b_1 x-\frac{b_1 \int D(t)dt}{a^2_1+b^2_1}$, and  $\delta_1=\frac{|\alpha_1| (a^2_1+b^2_1)^{3/4}}{2 a_1}$.  
From the above expression we have calculated the amplitude of the bright one soliton as  
\begin{align}\label{amplitude}
	A(t) = \sqrt{D(t)/R(t)} \sqrt{\frac{2(b_1+\sqrt{a_1^2+b_1^2})}{a_1^2+b_1^2}},
\end{align}
and the generalized velocity is $c(t)=\frac{\int D(t)dt}{a_1^2+b_1^2}$. One can see that the soliton envelope is explicitly influenced by the dispersion and nonlinearity coefficients $D(t)$ and $R(t)$. This is the significant difference from the bright one soliton solution of the constant coefficient FLE \cite{talukdar2023multi}. Unlike the NLSE case, where only the real part of the wave number contributes to the amplitude and imaginary part of the wave number contributes to the velocity of the soliton, here both the real and imaginary parts of the wave number influence the amplitude and, velocity, of the soliton solution. A typical bright one soliton solution of the vcFLE is depicted in Fig. \ref{fig:1ss}. 
%%%%%%%%%%%%%%%%%%%%%%%%%%%%%%%%%%%%%%%%%%%%%%%%%%%%%%%%%%%%%%%%%%%%%%%%%%%%%%%%%%%%%%%%%%%%%%%%%%%%%%%%%%%%%%%%%%%%%%%%%%%%%%%%%%%%%%%%%%%%%%%%%%%%%%%%%%%%%%%%%%%%%%%%%%%%%%%%%%%%%%%   
\subsection{Bright two soliton solution}
As explained above, similar procedure is followed to construct the two bright soliton solution of Eq. (\ref{model}). Now the initial solution is $G_1=\alpha_1 e^{\eta_1}+\alpha_2 e^{\eta_2}$, where $\eta_1=p_1 x+\Omega_1(t)$ and $\eta_2=p_2 x+\Omega_2(t)$. In this case the $G$ and $s$ series are truncated at the order $\epsilon^{3}$ and $F$ series is terminated at the order $\epsilon^{4}$. The auxiliary function is also expanded as $s= c_1 e^{\eta_1}+ c_2 e^{\eta_2} + c_3 e^{\eta_1+\eta^*_1+\eta_2} + c_4 e^{ \eta_1+\eta_2+\eta^*_2}$. After substituting the unknown functions in the form (\ref{bilin0}), the bright two soliton solution(2SS) of (\ref{model}) takes the form
\begin{align}
\label{sol3}
q(x,t)=\sqrt{\frac{D(t)}{R(t)}} \frac{\alpha_1 e^{\eta_1}+\alpha_2 e^{\eta_2}+\alpha_3 e^{\eta_1+\eta_1^*+\eta_2}+\alpha_4 e^{\eta_2+\eta_2^*+\eta_1}}{1 + \beta_1 e^{\eta_1+\eta_1^*}+ \beta_2 e^{\eta_2+\eta_2^*}+ \beta_3 e^{\eta_1+\eta_2^*}+ \beta_4 e^{\eta_1^*+\eta_2}+ \beta_5 e^{\eta_1+\eta_1^*+\eta_2+\eta_2^*}},
\end{align}
where
\begin{align}\label{par1}
	\Omega_1(t)&= \int \frac{D(t)dt}{p_1}, \quad 
	\Omega_2(t)=\int \frac{D(t)dt}{p_2},  \\ \label{par2}
	\alpha_3 &=\frac{i |\alpha_1|^2 \alpha_2 {p^*_1}^3 (p_1-p_2)^2}{(p_1+p^*_1)^2 (p^*_1+p_2)^2};
	\alpha_4 =\frac{i |\alpha_2|^2 \alpha_1 {p^*_2}^3 (p_1-p_2)^2}{ (p_2+p^*_2)^2 (p_1+p^*_2)^2},\\ \label{par3}
	\beta_1&=\frac{i |\alpha_1|^2 |p_1|^2 p_1}{ (p_1 + p^*_1)^2};
	\quad\quad\quad\hspace{0.5em}
	\beta_2=\frac{i |\alpha_2|^2 |p_2|^2 p_2}{ (p^*_2 + p_2)^2},\\ \label{par4}
	\beta_3&=\frac{i \alpha^*_2 \alpha_1 {p_1}^2 p^*_2}{ (p_1 + p^*_2)^2};
	\quad\quad\quad\hspace{0.5em}
	\beta_4=\frac{i \alpha_2 \alpha^*_1 p_2^2 p_1^*}{(p_2 + p^*_1)^2},\\ \label{par5}
	\beta_5&=\frac{- p_1 p_2 |p_1|^2 |p_2|^2 |\alpha_1|^2 |\alpha_2|^2  |(p_1-p_2)|^4}{ (p_1+p^*_1)^2 (p^*_1+p_2)^2 (p_1+p^*_2)^2 (p_2+p^*_2)^2}.
\end{align}    
Here $\alpha_1$, $\alpha_2$, $p_1=a_1+ib_1$, $p_2=a_2+ib_2$ are arbitrary complex parameters which actually control the properties of the given bright two soliton solution. We depict a typical bright 2SS in Fig. \ref{fig:2ss}.	
%%%%%%%%%%%%%%%%%%%%%%%%%%%%%%%%%%%%%%%%%%%%%%%%%%%%%%%%%%%%%%%%%%%%%%%%%%%%%%%%%%%%%%%%%%%%%%%%%%%%%%%%%%%%%%%%%%%%%%%%%%%%%%%%%%%%%%%%%%%%%%%%%%%%%%%%%%%%%%%%%%%%%%%%%%%%%%%%%
\subsection{$N$-bright soliton solution}
Generalizing the above procedure using the initial seed solution as $G_{1}=\sum_{n=1}^{N}\alpha_{n}e^{\eta_{n}}$, one can get the $N$- bright soliton solution of (\ref{model}) which can be written in the form,
\begin{align}
\label{Nsol}
q(x,t)=\sqrt{\frac{D(t)}{R(t)}} \frac{ \sum_{n=1}^{N} \epsilon^{2n-1} G_{2n-1}}{1+\sum_{n=1}^{N} \epsilon^{2n} F_{2n}},\\ \label{Nsol2}
q^*(x,t)=\sqrt{\frac{D(t)}{R(t)}} \frac{ \sum_{n=1}^{N} \epsilon^{2n-1} G^*_{2n-1}}{1+\sum_{n=1}^{N} \epsilon^{2n} F^*_{2n}}.	
\end{align}
Proceeding as in the cases of $N=1$ and $N=2$ above, we can obtain the arbitrary $N$-soliton solution systematically. 
%%%%%%%%%%%%%%%%%%%%%%%%%%%%%%%%%%%%%%%%%%%%%%%%%%%%%%%%%%%%%%%%%%%%%%%%%%%%%%%%%%%%%%%%%%%%%%%%%%%%%%%%%%%%%%%%%%%%%%%%%%%%%%%%%%%%%%%%%%%%%%%%%%%%%%%%%%%%%%%%%%%%%%%%%%%%%%%%%
\section{Interaction dynamics of bright solitons}
In this section we present the collision dynamics of the bright soliton solutions of the vcFLE. To understand the collision behaviour one has to perform a systematic asymptotic analysis for the two soliton solution \cite{mlbook}. Here we have deduced the asymptotic forms of the two soliton solution (\ref{sol3}) as $t \rightarrow \pm \infty$. These expressions will give the exact analytical forms of the solitons before and after collision. We consider that before the collision the two bright solitons $S_{1}$ and $S_{2}$ are well separated and then they undergo collision process as they move with the relative velocity.
\subsection{Asymptotic analysis}
\begin{figure*}[]
	\centering
	\subfloat[]{%
		\includegraphics[width=0.45\textwidth,height=0.2\textheight]{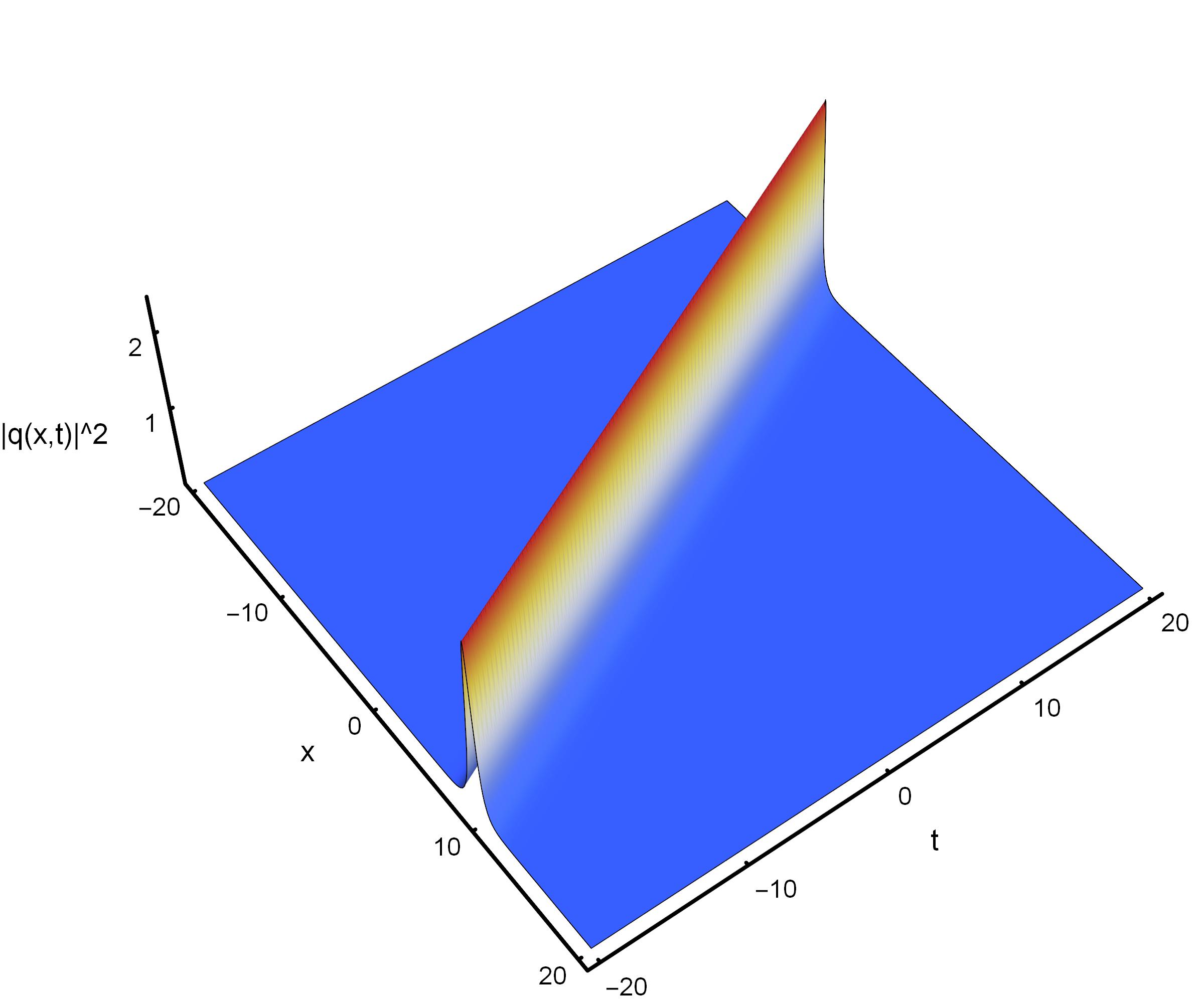}%
		\label{fig:1ss}%
	}\hfill
	\subfloat[]{%
		\includegraphics[width=0.45\textwidth,height=0.2\textheight]{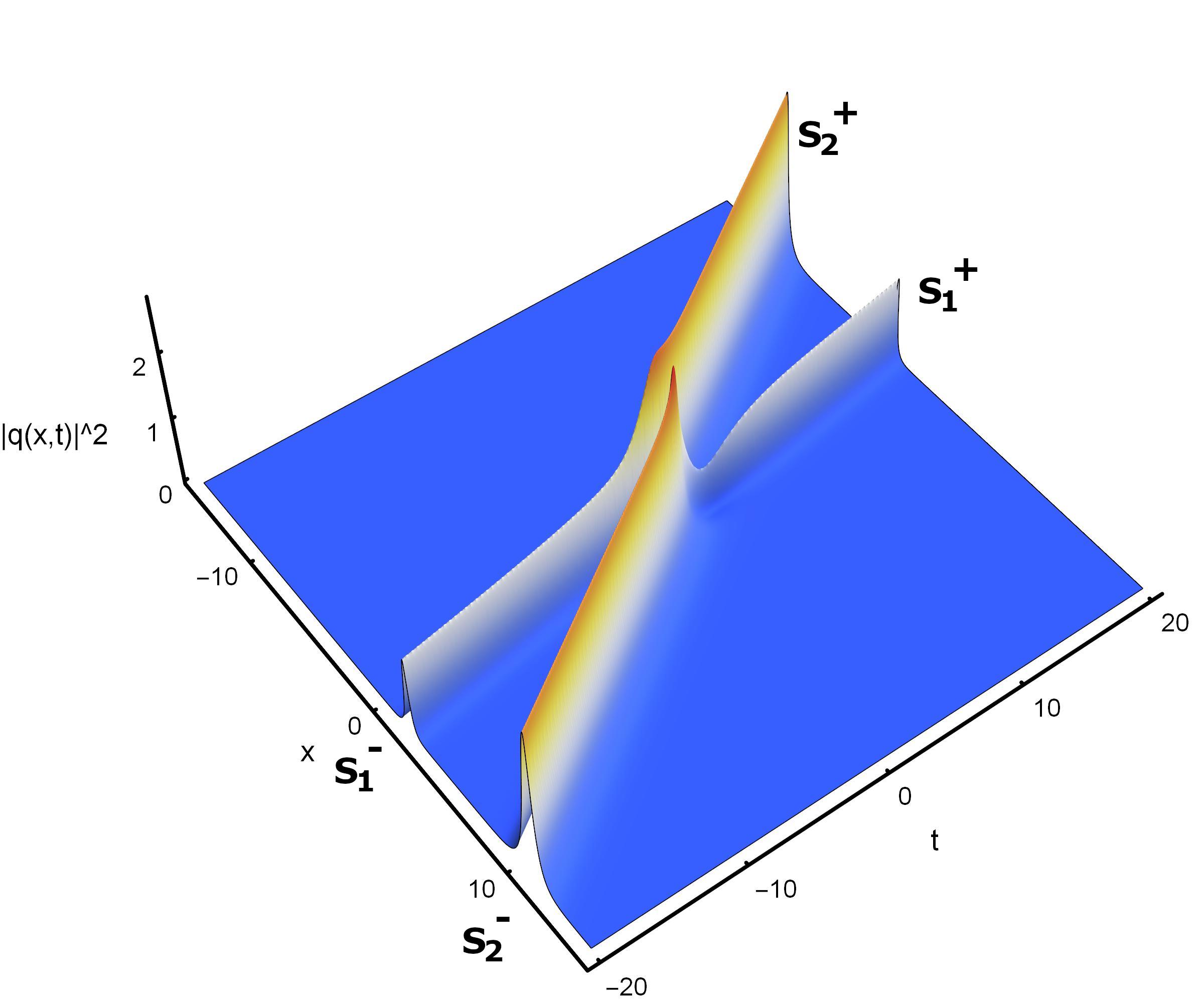}%
		\label{fig:2ss}%
	}
	
	\caption{\scriptsize (a) is the 3D plot of 1SS with $p_1=1+i$, $\alpha_1=5+0.5i$.
		(b) is the 3D plot of 2SS with $p_1=1+i$, $p_2=2+2i$, $\alpha_1=5+0.5i$, and $\alpha_2=-5-0.7i$.}
	\label{fig}
\end{figure*}

A. Before collision ($t \rightarrow -\infty$):\\
In the limit $t \rightarrow -\infty$, the solution (\ref{sol3}) can be reduced to the following forms:\\
\underline{Soliton 1}: $\eta_{1R} \to 0$, $\quad \eta_{2R} \to -\infty$,
\begin{align}
	S_1|^-&=\sqrt{\frac{D(t)}{R(t)}} \frac{\alpha_1 e^{\eta_1}}{1 + \beta_1 e^{\eta_1+\eta_1^*}}, \nonumber\\ \label{s1-}
	S_1|^-&=\sqrt{\frac{D(t)}{R(t)}} 
	\frac{\alpha_1}{2} e^{\frac{\eta_1-\eta_1^*-ln\beta_1}{2}} \sech \bigg(\frac{\eta_1+\eta_1^*+ln\beta_1}{2}\bigg),	
\end{align}
\underline{Soliton 2}: $\eta_{2R} \to 0$, $\quad \eta_{1R} \to \infty$, 	 
\begin{align}	
	S_2|^-&=\sqrt{\frac{D(t)}{R(t)}} \frac{\alpha_3 e^{\eta_2}}{\beta_1 + \beta_5 e^{\eta_2+\eta_2^*}}, \nonumber\\ \label{s2-}
	S_2|^-&=\sqrt{\frac{D(t)}{R(t)}} 
	\frac{\alpha_3}{2\beta_1} e^{\frac{\eta_2-\eta_2^*-ln\frac{\beta_5}{\beta_1}}{2}} \sech \bigg(\frac{\eta_2+\eta_2^*+ln\frac{\beta_5}{\beta_1}}{2}\bigg), 	  
\end{align}
Here $S_{1}|^{-}$ and $S_{2}|^{-}$ indicate the solitons $S_{1}$ and $S_{2}$ before the collision.\\ 
B. After collision ($t \to +\infty$):\\
In the limit $t \rightarrow +\infty$, the solution (\ref{sol3}) can be reduced to the following forms:\\
\underline{Soliton 1}: $\eta_{1R} \to 0$, $\quad \eta_{2R} \to \infty$, 
\begin{align}
	S_1|^+&=\sqrt{\frac{D(t)}{R(t)}} \frac{\alpha_4 e^{\eta_1}}{\beta_2 + \beta_5 e^{\eta_1+\eta_1^*}}, \nonumber\\ \label{s1+}
	S_1|^+&=\sqrt{\frac{D(t)}{R(t)}} 
	\frac{\alpha_4}{2\beta_2} e^{\frac{\eta_1-\eta_1^*-ln\frac{\beta_5}{\beta_2}}{2}} \sech \bigg(\frac{\eta_1+\eta_1^*+ln\frac{\beta_5}{\beta_2}}{2}\bigg),
\end{align}	
\underline{Soliton 2}: $\eta_{2R} \to 0$, $\quad \eta_{1R} \to -\infty$,
\begin{align}	
	S_2|^+&=\sqrt{\frac{D(t)}{R(t)}} \frac{\alpha_2 e^{\eta_2}}{1 + \beta_2 e^{\eta_2+\eta_2^*}}, \nonumber\\ \label{s2+}
	S_2|^+&=\sqrt{\frac{D(t)}{R(t)}} 
	\frac{\alpha_2}{2} e^{\frac{\eta_2-\eta_2^*-ln\beta_2}{2}} \sech \bigg(\frac{\eta_2+\eta_2^*+ln\beta_2}{2}\bigg).
\end{align}
Here $S_{1}|^{+}$ and $S_{2}|^{+}$ indicate the solitons $S_{1}$ and $S_{2}$ after the collision.\\
\indent  The parameters $\alpha_1$, $\alpha_2$ are arbitrary complex constants and $\alpha_3$, $\alpha_4$, $\beta_1$, $\beta_2$, $\beta_5$ are defined earlier in Eqs. (\ref{par2}), (\ref{par3}), and (\ref{par5}) respectively. The amplitudes of solitons $S_1$ and $S_2$ given by Eqs. (\ref{s1-}), (\ref{s1+}) and Eqs. (\ref{s2-}), (\ref{s2+}) are same which means physically before and after collision the amplitude of the individual solitons are same. Let $A_j|^-$, $j=1,2,$ be the amplitudes of $S_1$ and $S_2$ before collision and $A_j|^+$ be their amplitudes after collision. Then
\begin{align}\label{amp}
	A_j|^-=A_j|^+.
\end{align} 
The amplitudes of the solitons can be extracted from modulus square of the solution as expressed in Eq. (\ref{amplitude}). So, we calculate a transition amplitude $T_j=\frac{A_j|^+}{A_j|^-}$, $j=1,2$, which relates the amplitudes of solitons $S_1$, $S_2$ (after and before collision) given by Eqs. (\ref{s1-}-\ref{s2+}). In accordance with Eq. (\ref{amp}) we get, $T_j=1$. In this case, one obtains the shape preserving elastic collisions of the solitons as seen in Fig. \ref{fig:2ss}. Subsequently, one can also obtain the transition intensity as $|T_j|^2=1$. \\
\indent Besides the transition amplitude and intensity, there is another important quantity associated with the collision process i.e. phase shift. For soliton $S_1$ given by Eqs. (\ref{s1-}), (\ref{s1+}), the initial phase is $\frac{1}{2}ln\beta_1$ which gets changed to $\frac{1}{2}ln\frac{\beta_5}{\beta_2}$ after collision. Hence the corresponding phase shift is,
\begin{align}\nonumber
	\Phi_1&=\frac{1}{2}(ln\frac{\beta_5}{\beta_2}-ln\beta_1), \\
	&=\frac{1}{2}ln\frac{\beta_5}{\beta_1\beta_2}=\frac{1}{2}ln\frac{\left((a_1-a_2)^2+(b_1-b_2)^2\right)^2}{\left((a_1+a_2)^2+(b_1-b_2)^2\right)^2}.
\end{align}
Proceeding with the same technique, the phase shift for soliton $S_2$ given by Eqs. (\ref{s2-}), (\ref{s2+}) is,
\begin{align}\nonumber
	\Phi_2&=\frac{1}{2}(ln\beta_2-ln\frac{\beta_5}{\beta_1}), \\ \nonumber
	&=\frac{1}{2}ln\frac{\beta_1\beta_2}{\beta_5}, \\
	&=-\frac{1}{2}ln\frac{\beta_5}{\beta_1\beta_2}=-\Phi_1.
\end{align} 
\indent Though the two individual bright solitons undergo shape preserving elastic collision there is a small phase shift occurred during the collision process. This is a typical soliton collision behaviour. Further it is important to notice that, the phase shift depends only on the soliton parameters $a_j$, and $b_j$, $j=1,2$. The other complex constants $\alpha_1$, and $\alpha_2$ do not affect the phase shift process. Due to this phase shift, the relative separation distance between the two solitons also vary during the collision process. This can be calculated by the difference between the separation of $S_1$ and $S_2$ after collision and before collision which is given by
\begin{align}
	\Delta x_{12}=x^+_{12}-x^-_{12}=\frac{a_1+a_2}{a_1a_2}\Phi_1.
\end{align}           
\indent One can easily notice that the above two quantities, namely phase shift and relative separation distance, depend explicitly on the two soliton parameters $a_j$ and $b_j$ which in turn also determine the amplitudes of the solitons. Meanwhile, the dispersion and nonlinearity coefficients $D(t)$ and $R(t)$ do not enter into the expressions for the above two quantities.

%%%%%%%%%%%%%%%%%%%%%%%%%%%%%%%%%%%%%%%%%%%%%%%%%%%%%%%%%%%%%%%%%%%%%%%%%%%%%%%%%%%%%%%%%%%%%%%%%%%%%%%%%%%%%%%%%%%%%%%%%%%%%%%%%%%%%%%%%%%%%%%%%%%%%%%%%%%%%%%%%%%%%%%
\subsection{Accelerating/Retarding solitons with zero gain/loss}
To study the accelerating or retarding soliton we
consider the dispersion coefficient $D(t)$ and nonlinearity coefficient $R(t)$ to be equal so that the gain/loss parameter $\Gamma(t)=0$, which ensures no loss/gain in the amplitude of the soliton. In this case, the soliton now has a time dependent velocity, and can be accelerated without changing the soliton width i.e. keeping $a_1$ constant. Here one can always accelerate the obtained soliton envelope in Eq. (\ref{q1}) by manipulating $b_{1}$ and $D(t)$.
\begin{figure*}[]
	\centering
	\subfloat[]{%
		\includegraphics[width=0.40\textwidth]{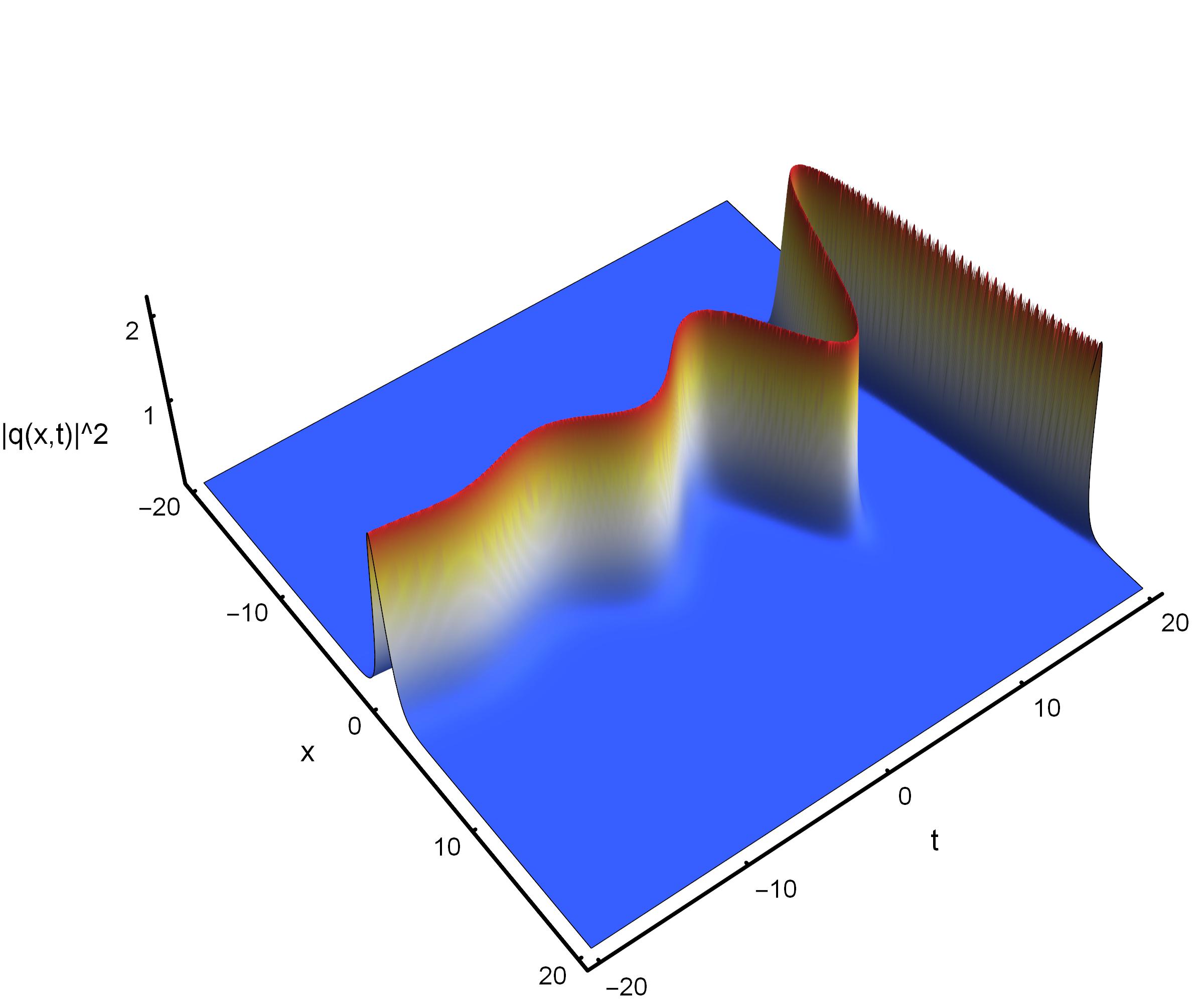}%
		\label{fig:2a}%
	}%\hfill
	\subfloat[]{%
		\includegraphics[width=0.40\textwidth]{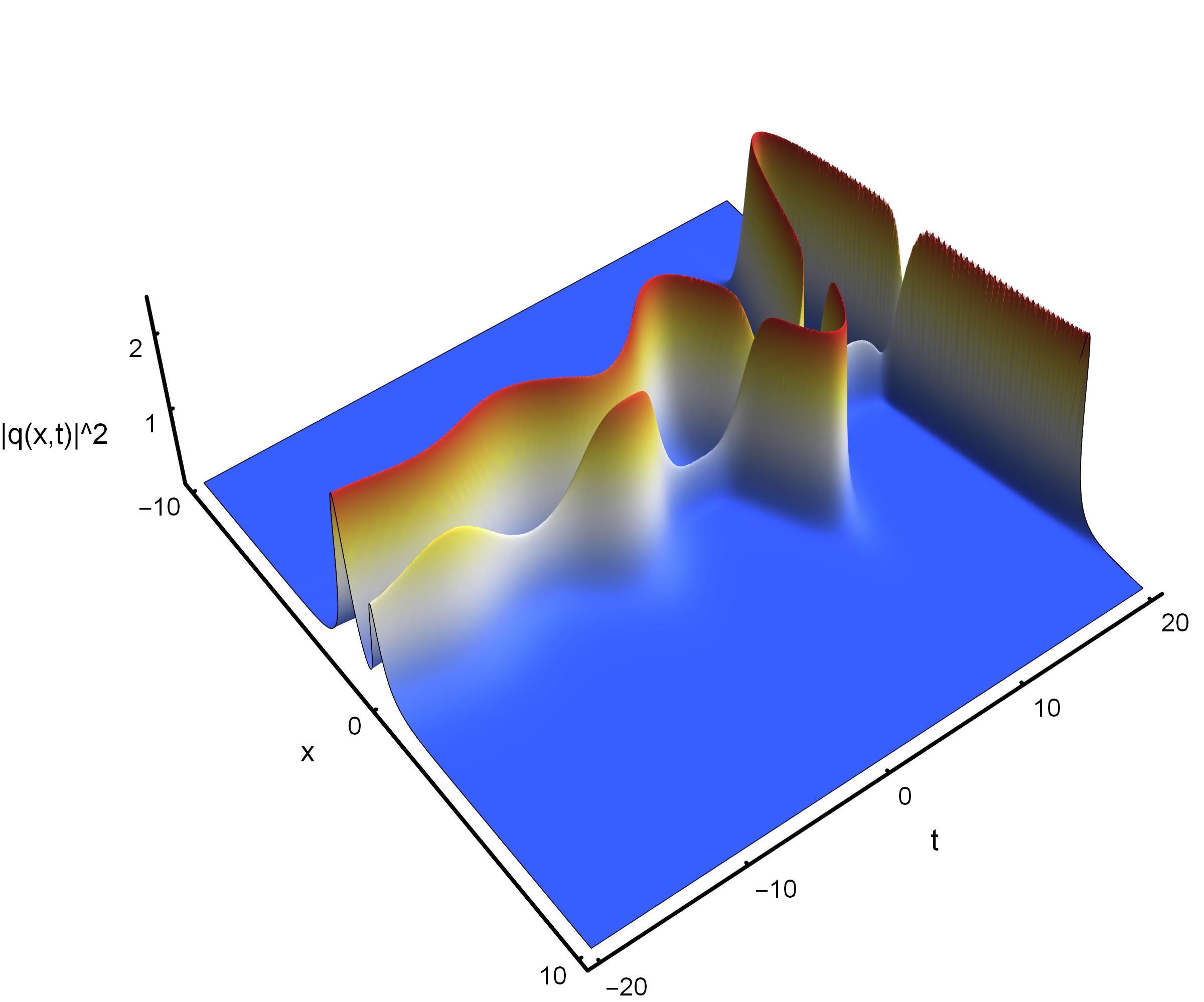}%
		\label{fig:2b}%
	}\\[4pt]
	\subfloat[]{%
		\includegraphics[width=0.40\textwidth]{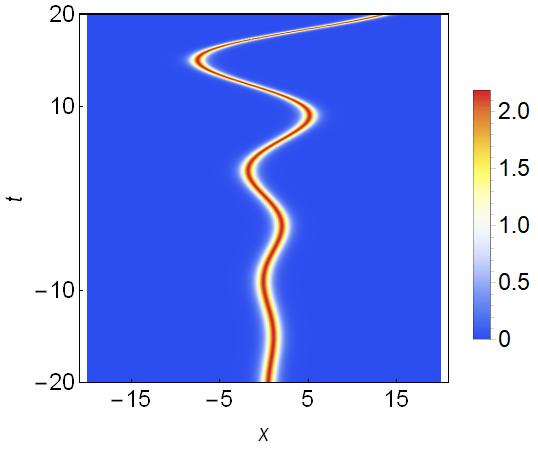}%
		\label{fig:2c}%
	}%\hfill
	\subfloat[]{%
		\includegraphics[width=0.40\textwidth]{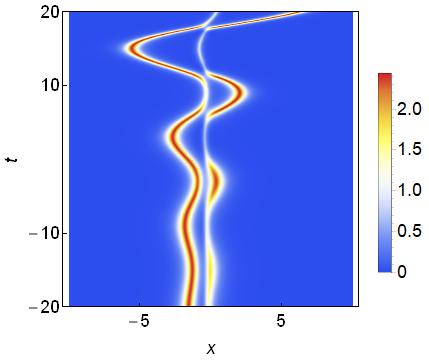}%
		\label{fig:2d}%
	}
	
	\caption{\scriptsize (a) shows the periodically accelerated/retarded 1SS when
		$D(t)=e^{\sigma t}\cos(\phi t)$, $R(t)=e^{\sigma t}\cos(\phi t)$, $\Gamma(t)=0$, where
		$\sigma=0.1$, $\phi=\pi/6$, with $p_1=1+0.1i$, $\alpha_1=1+0.1i$.
		(b) shows the elastic interaction between 2SS when
		$D(t)=e^{\sigma t}\cos(\phi t)$, $R(t)=e^{\sigma t}\cos(\phi t)$, $\Gamma(t)=0$, where
		$\sigma=0.1$, $\phi=\pi/6$, with $p_1=1+i$, $p_2=3+2i$, $\alpha_1=5+0.5i$, $\alpha_2=-5-0.5i$.
		The corresponding density plots are shown in (c), (d) respectively under the same parametric values.}
	\label{fig:2}
\end{figure*}
For this case, we have chosen $D(t)$, and $R(t)$ as two linearly dependent functions as follows,
\begin{align}
	D(t)&=e^{\sigma t}\cos(\phi t),\label{dt}\\
	R(t)&=e^{\sigma t}\cos(\phi t), \\
	\Gamma(t)&=0.
\end{align} 
The dispersion coefficient present in the velocity term $c(t)$ influences the propagation i.e. the velocity is modulated by the term $D(t)$. The specific function $D(t)$, as defined in $(\ref{dt})$, explicitly leads to $c(t)=\frac{e^{\sigma t}(\sigma \cos \phi t+\phi \sin \phi t)}{(a_{1}^{2}+b_{1}^{2})(\sigma^{2}+\phi^{2})}$. In this scenario the soliton is accelerated/retarded periodically. The $\cos(\phi t)$ term gives the periodicity in acceleration/retardation while the exponential term $e^{\sigma t}$ is responsible for the rapid acceleration/retardation as $t$ goes from (-20 to 20) as seen in Figs. \ref{fig:2a}-\ref{fig:2d} for both 1SS and 2SS.
%%%%%%%%%%%%%%%%%%%%%%%%%%%%%%%%%%%%%%%%%%%%%%%%%%%%%%%%%%%%%%%%%%%%%%%%%%%%%%%%%%%%%%%%%%%%%%%%%%%%%%%%%%%%%%%%%%%%%%%%%%%%%%%%%%%%%%%%%%%%%%%%%%%%%%%%%%%%%%%%%%%%%%%%%%%%%%%%%%%%%%% 
\subsection{Soliton tunneling through dispersion and nonlinearity barrier}
\begin{figure*}[]
	\centering
	\subfloat[]{%
		\includegraphics[width=0.40\textwidth]{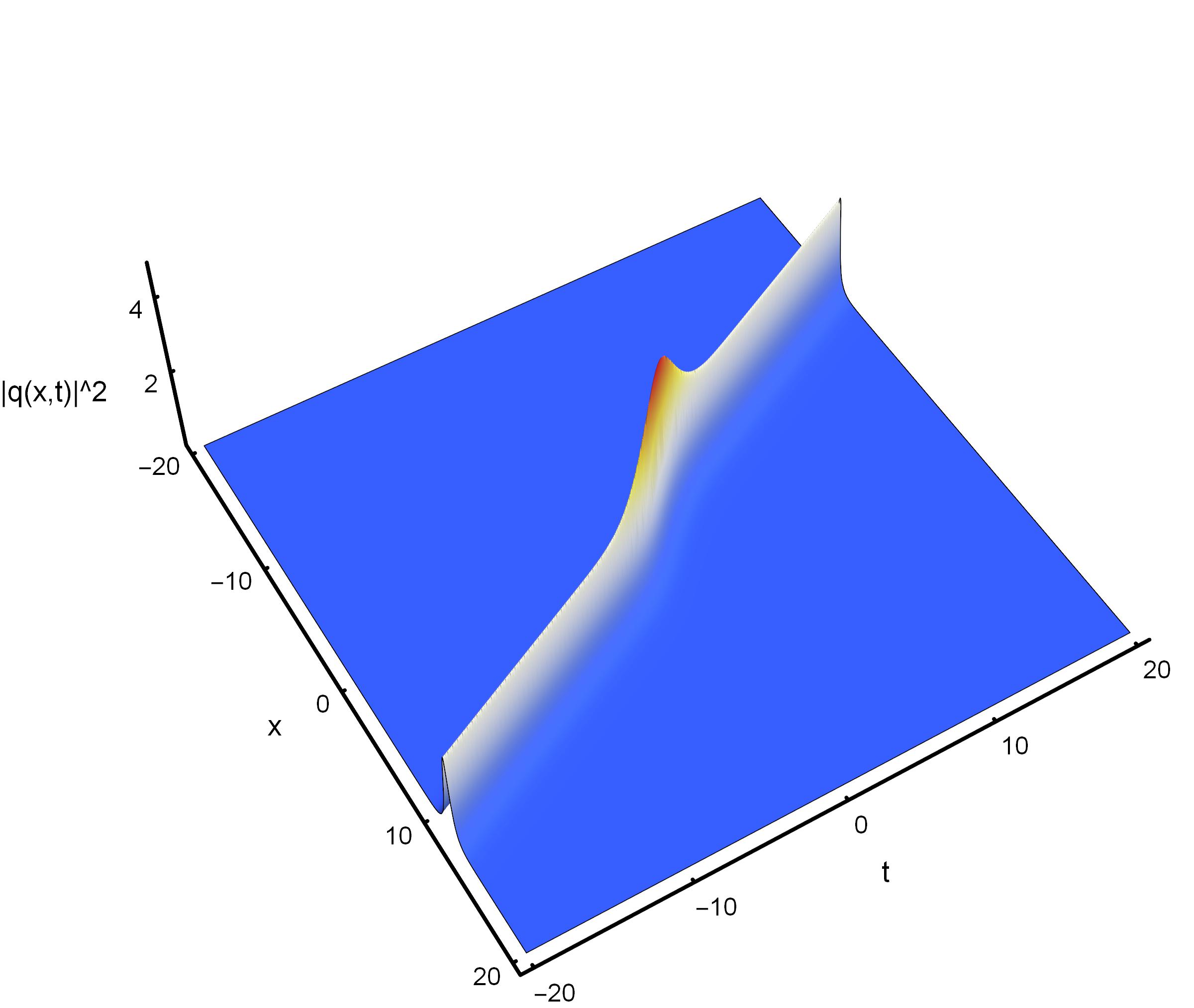}%
		\label{fig:3a}%
	}%\hfill
	\subfloat[]{%
		\includegraphics[width=0.40\textwidth]{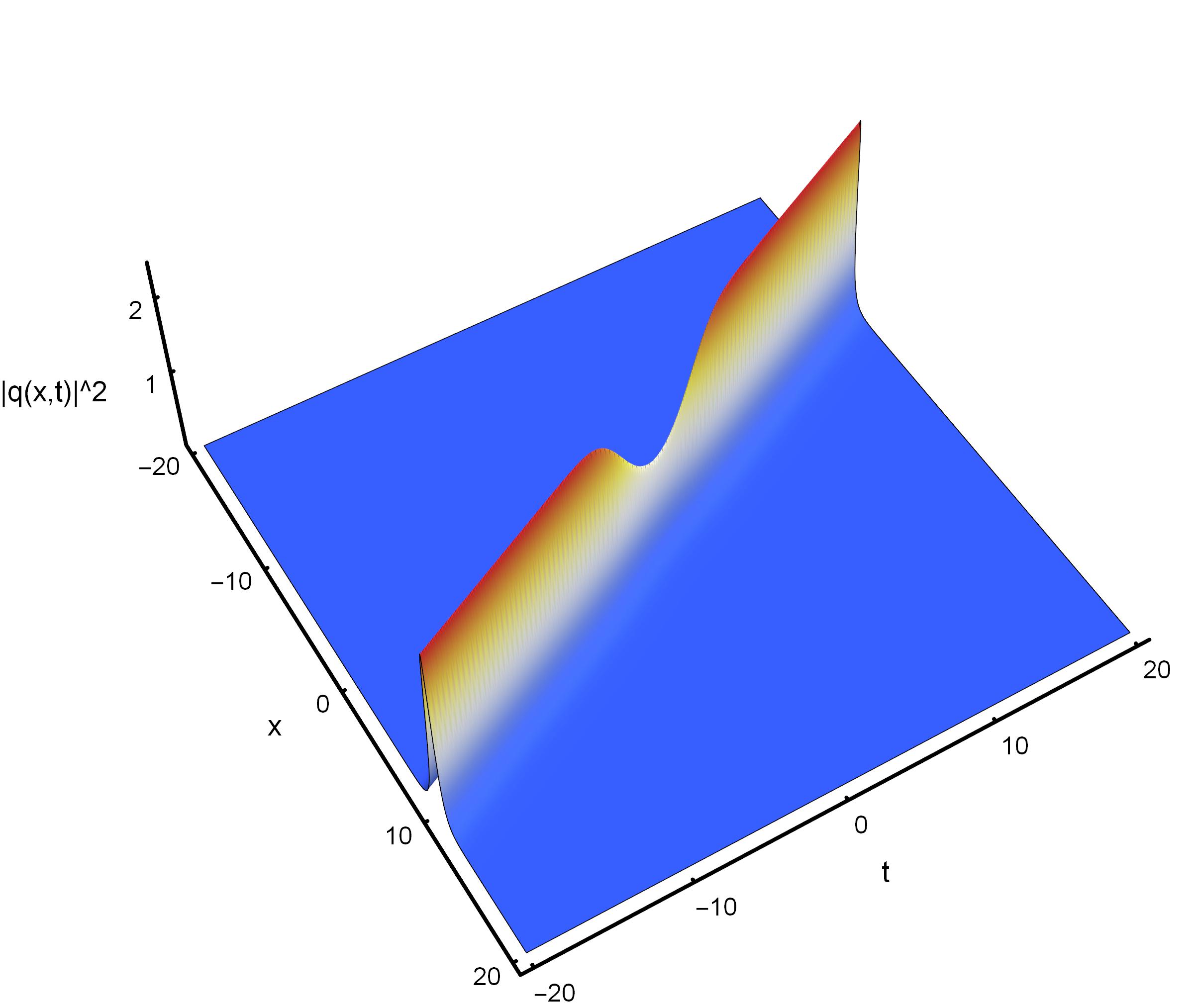}%
		\label{fig:3b}%
	}\\[4pt]
	\subfloat[]{%
		\includegraphics[width=0.40\textwidth]{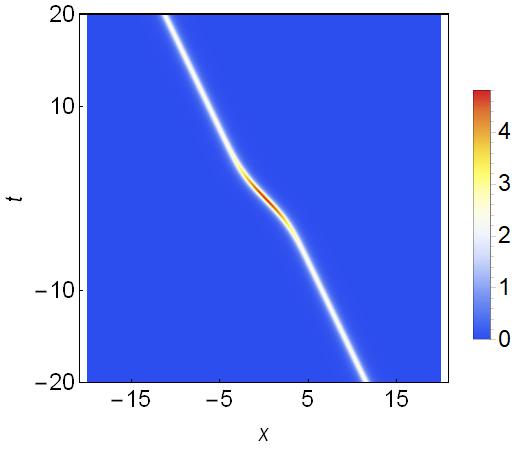}%
		\label{fig:3c}%
	}%\hfill
	\subfloat[]{%
		\includegraphics[width=0.40\textwidth]{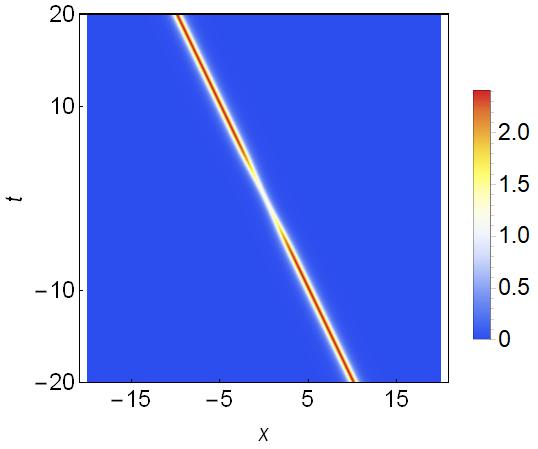}%
		\label{fig:3d}%
	}
	
	\caption{\scriptsize The 3D plot of (a) a 1SS tunneling through the dispersion barrier
		$D(t)=1+e^{-0.1t^2}$, $R(t)=1$, and $\Gamma(t)=-\frac{0.1 t e^{-0.1t^2}}{1+ e^{-0.1 t^2}}$,
		with $p_1=1+i$, $\alpha_1=1+2i$.
		In (b) we show the tunneling through nonlinear barrier where
		$R(t)=1+e^{-0.1t^2}$, $D(t)=1$, and
		$\Gamma(t)=\frac{0.1 t e^{-0.1t^2}}{1+ e^{-0.1 t^2}}$, $p_1=1+i$, $\alpha_1=1+2i$.
		Plots (c), (d) show the density distributions of the corresponding 3D plots (a), (b) respectively
		under the same parametric values.}
	\label{fig:3}
\end{figure*}

\begin{figure*}[]
	\centering
	\subfloat[]{%
		\includegraphics[width=0.40\textwidth]{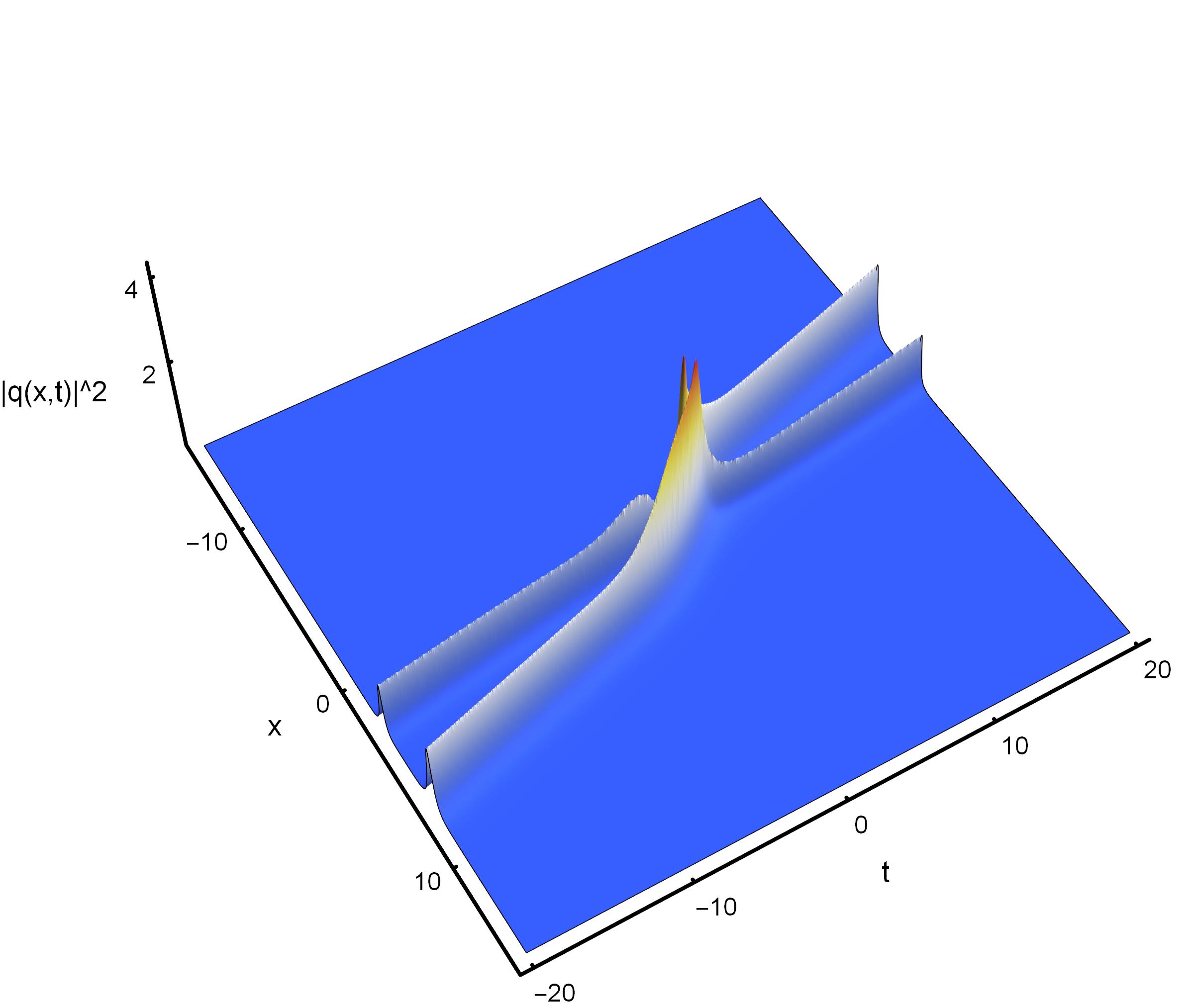}%
		\label{fig:4a}%
	}%\hfill
	\subfloat[]{%
		\includegraphics[width=0.40\textwidth]{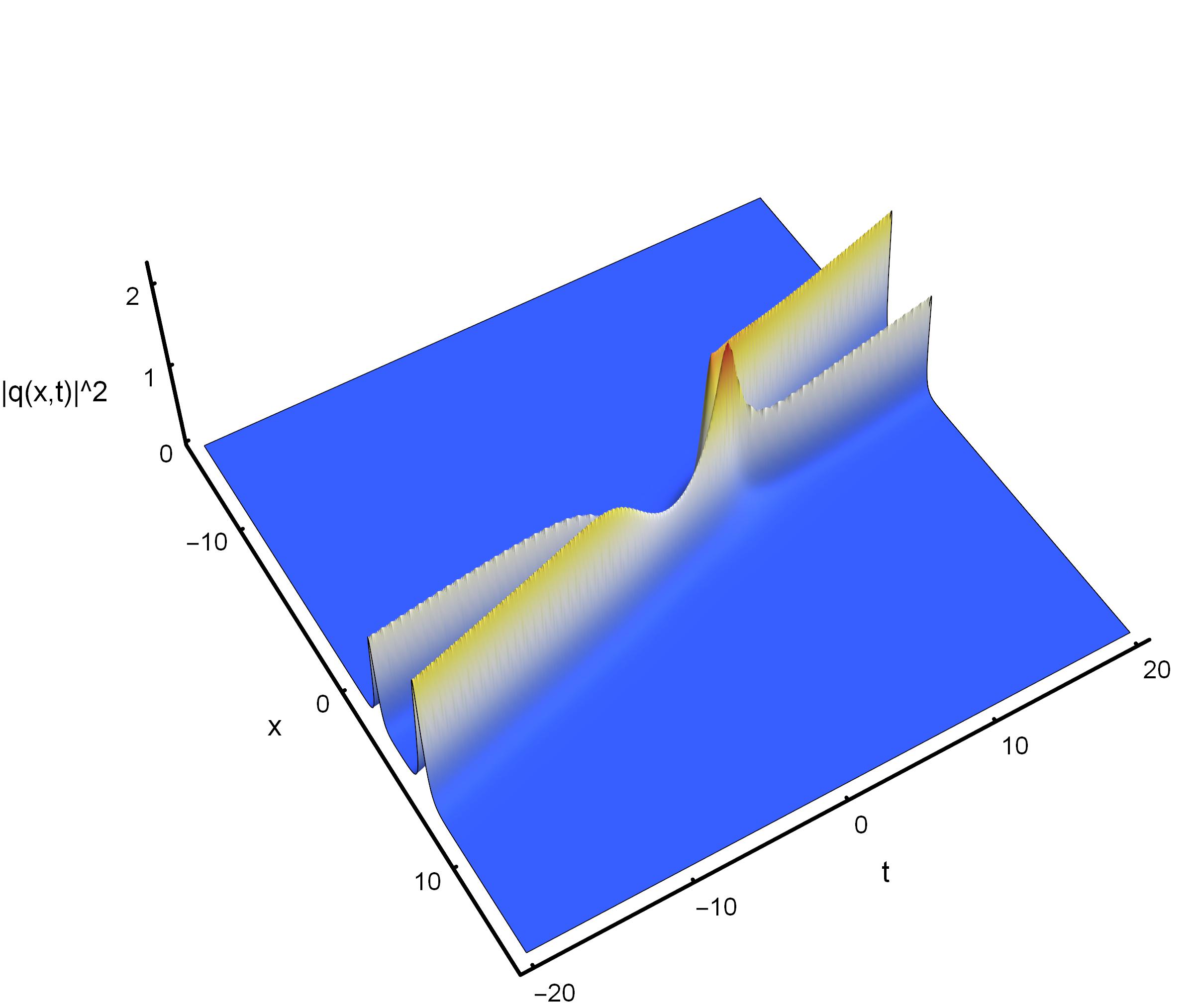}%
		\label{fig:4b}%
	}\\[4pt]
	\subfloat[]{%
		\includegraphics[width=0.40\textwidth]{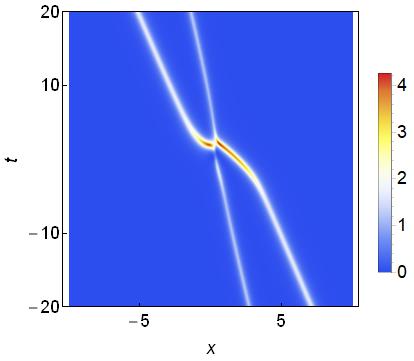}%
		\label{fig:4c}%
	}%\hfill
	\subfloat[]{%
		\includegraphics[width=0.40\textwidth]{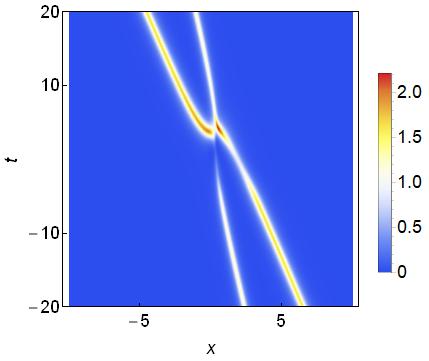}%
		\label{fig:4d}%
	}
	
	\caption{\scriptsize The 3D plot of (a) 2SS tunneling through the dispersion barrier
		$D(t)=1+e^{-0.1t^2}$, $R(t)=1$, and
		$\Gamma(t)=-\frac{0.1 t e^{-0.1 t^2}}{e^{-0.1 t^2}+1}$, with
		$p_1=1.5+1.5i$, $p_2=2+2i$, $\alpha_1=1+0.5i$, $\alpha_2=-1-0.5i$.
		In (b) we show the tunneling through the nonlinear barrier
		$R(t)=1+e^{-0.1t^2}$, $D(t)=1$, and
		$\Gamma(t)=\frac{0.1 t e^{-0.1 t^2}}{e^{-0.1 t^2}+1}$, with
		$p_1=1.5+1.5i$, $p_2=2+2i$, $\alpha_1=1+0.5i$, $\alpha_2=-1-0.5i$.
		Plots (c), (d) show the density distribution of the corresponding 3D plots (a), (b) respectively
		under the same parametric values.}
	\label{fig:4}
\end{figure*}

\indent In this subsection we want to discuss the nonlinear tunneling effect of the vcFLE bright solitons. The phenomenon of nonlinear tunneling based on the choice of dispersion and nonlinearity has been already studied for various systems in refs.  \cite{serkin2001nonlinear,yang2006cascade,zhong2010soliton,newell1978nonlinear,dai2012controllable,dai2010nonlinear,serkin2013geiger,he2011self,mani2013dispersion, musammil2017ultrashort, nandy2021dark}. All these works have shown that for various physical dispersion or nonlinearity coefficients the solitons can pass through the barrier/well. To investigate the nonlinear tunneling of vcFLE bright solitons through dispersion barrier we consider the dispersion coefficient as a Gaussian function as follows,
\begin{align}\label{barrier1}
	D(t)&=\Delta_0+\Delta_1 e^{-\Delta_2 (t-t_0)^2},\\
	R(t)&=1,\\ \label{barrier1.1}
	\Gamma(t)&=-\frac{\Delta_1 \Delta_2 (t-t_0) e^{-\Delta_2(t-t_0)^2}}{\Delta_0+\Delta_1 e^{-\Delta_2 (t-t_0)^2}},
\end{align}
where $\Delta_0$ is the constant dispersion and $\Delta_1$, $\Delta_2$ are the barrier height and width respectively with respect to the soliton amplitude. $t_0$ represents the coordinate that determines location of the barrier. Similarly, if we choose the same Gaussian function as the nonlinearity coefficient, it gives us the nonlinearity barrier as follows,
\begin{align}
	D(t)&=1,\\ \label{barrier2}
	R(t)&=\Delta_0+\Delta_1 e^{-\Delta_2 (t-t_0)^2},\\
	\Gamma(t)&=\frac{\Delta_1 \Delta_2 (t-t_0)e^{-\Delta_2(t-t_0)^2}}{\Delta_0+\Delta_1 e^{-\Delta_2 (t-t_0)^2}}.
\end{align}        
\indent For both the dispersion as well as nonlinearity barrier, we have considered $D(t)$ and $R(t)$ to be linearly independent that ensures $\Gamma(t)$ to vanish at the barrier location $t=t_0$ and remains non-vanishing otherwise. \\ 
\indent  Now, when the bright soliton propagates through the dispersion barrier given by Eq. (\ref{barrier1}), the amplitude/intensity of the soliton is diminished as the function $\Gamma(t)$ for this case given by Eq. (\ref{barrier1.1}) is negative, which corresponds to the loss in the amplitude/intensity. However, at the barrier location $t=t_{0}$ the function $\Gamma(t)$ becomes zero, which indicates no loss, and we can see the formation of a mountain (peak). On the other hand, when the soliton propagates through the nonlinearity barrier given by Eq. (\ref{barrier2}), the function $\Gamma(t)$ indicates a gain in the amplitude as it is positive. Meanwhile, at the barrier location $t=t_{0}$, the amplitude/intensity drops and a valley is formed, as evident from Fig. \ref{fig:3}. After tunneling through both the mountain and valley the soliton preserves its original shape. The nonlinear tunneling of 1SS and 2SS is shown in Figs. \ref{fig:3} and \ref{fig:4} respectively.\\ 
\indent In Figs. \ref{fig:3a} and \ref{fig:3c}, we consider $D(t)$ and $R(t)$ to be  $D(t)=1+e^{-0.1t^2}$ and $R(t)=1$ and $\Gamma(t)=-\frac{0.1 te^{-0.1t^2}}{1+ e^{-0.1 t^2}}$ i.e. $\Gamma(t)$ is non-vanishing and a function of time. As a result, the soliton experiences a dispersion barrier of $D(t)=1+e^{-0.1t^2}$ and in this case i.e. from the first term, the soliton receives a constant dispersion of $1$ and the second term is responsible for providing a Gaussian dispersion barrier for the soliton with a certain barrier of height $1$ and a width of $0.1$ with respect to the amplitude. It is evident from Figs. \ref{fig:3a} and \ref{fig:3c} that far from $t=0$, where $D(t)\rightarrow1$, the velocity varies with respect to $a_1$ and $b_1$, but when the soliton approaches $t=0$ region, the velocity is modulated by the Gaussian barrier dispersion $e^{-0.1t^2}$. One can see a rapid accumulation of energy at $t=0$. We can also verify this by looking at the loss parameter $\Gamma(t)$. In the region far from $t=0$, there is a net loss in the soliton amplitude $A(t)$ as $\Gamma(t)$ is negative in this region. But at $t=0$, $\Gamma(t)$ goes to zero and there is no net loss in amplitude. Thus, we can say in this scenario, the dispersion function produces a localized accumulation of energy around the region $t=0$ and the soliton tunnels through the dispersion barrier. We use the word tunnel/tunneling since it is analogous to tunneling effect in quantum mechanics. Similar, phenomenon is also seen in the case of 2SS as shown in Figs. \ref{fig:4a} and \ref{fig:4c}.\\
\begin{figure}[]
	\centering
	\subfloat[]{%
		\includegraphics[width=0.40\textwidth]{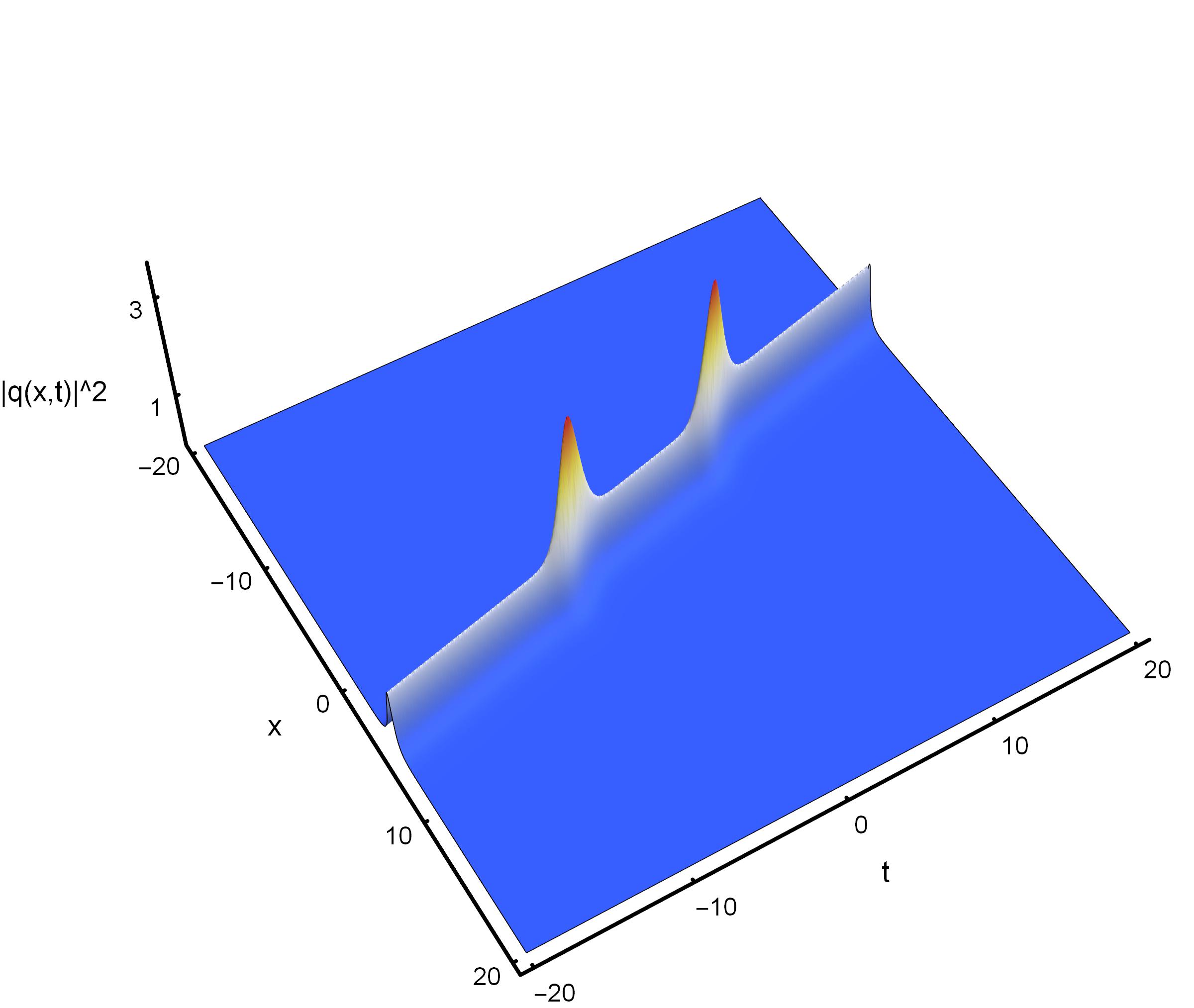}%
		\label{fig:5a}%
	}%\hfill
	\subfloat[]{%
		\includegraphics[width=0.40\textwidth]{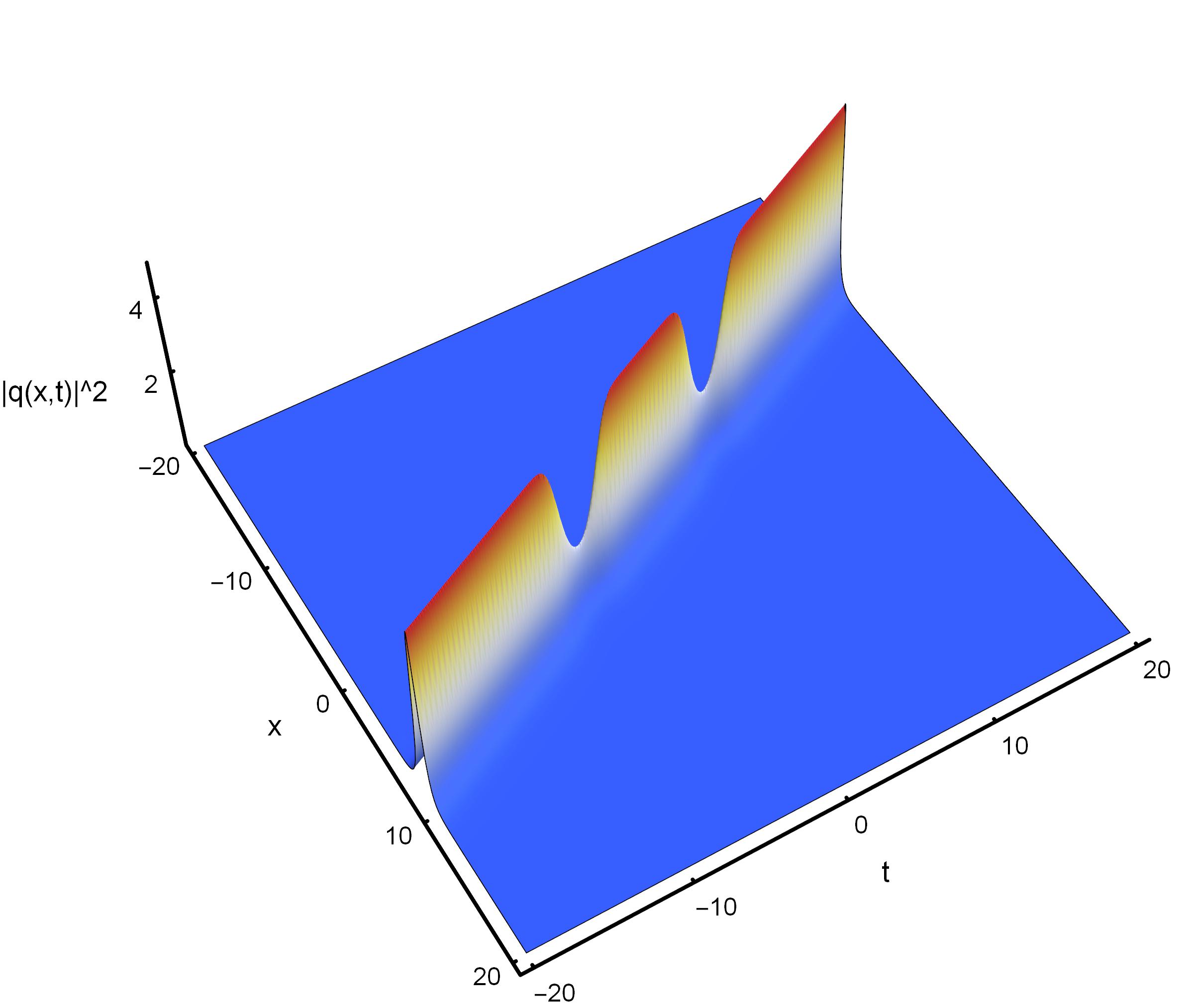}%
		\label{fig:5b}%
	}\\[4pt]	
	\subfloat[]{%
		\includegraphics[width=0.40\textwidth]{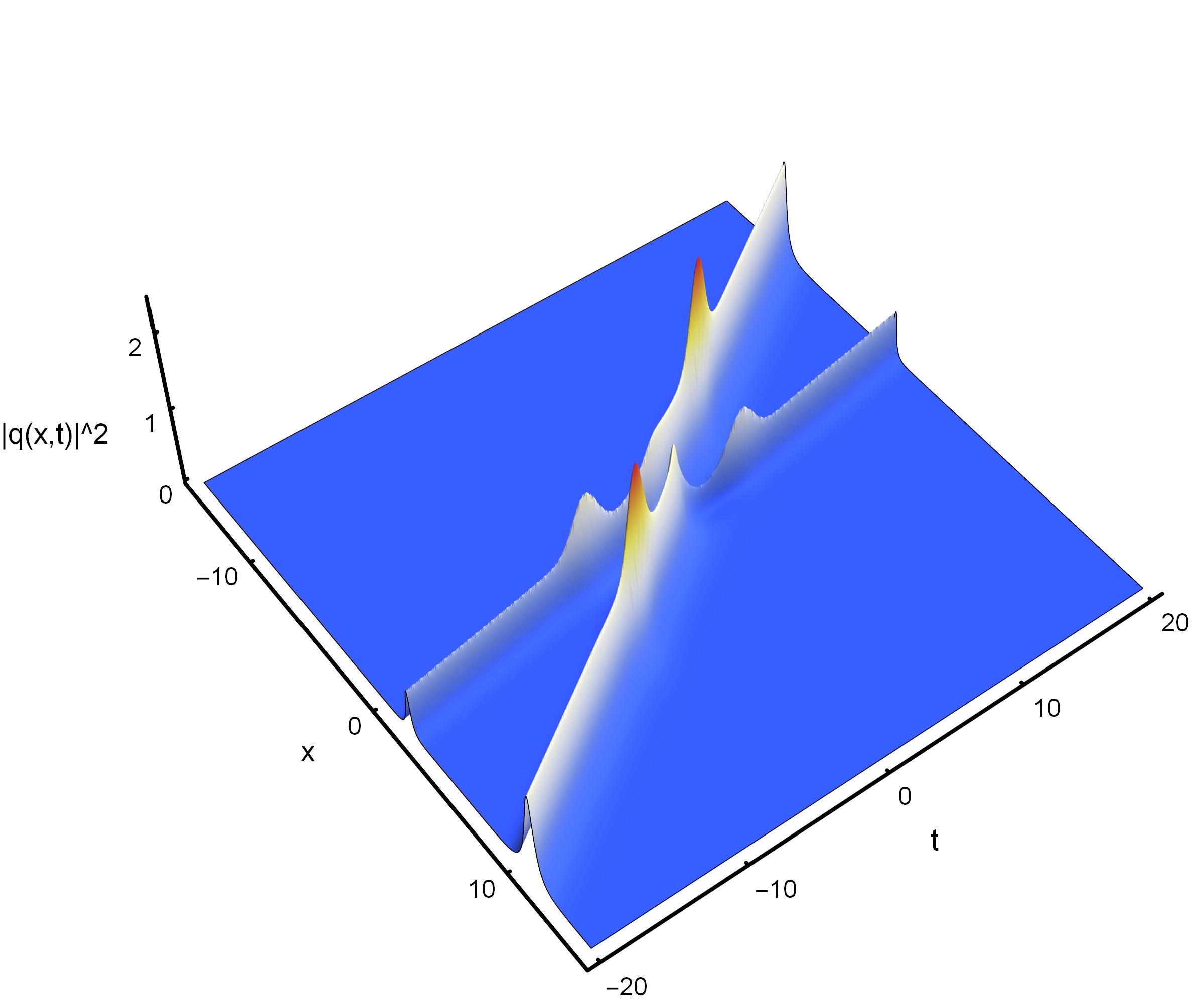}%
		\label{fig:5c}%
	}%\hfill
	\subfloat[]{%
		\includegraphics[width=0.40\textwidth]{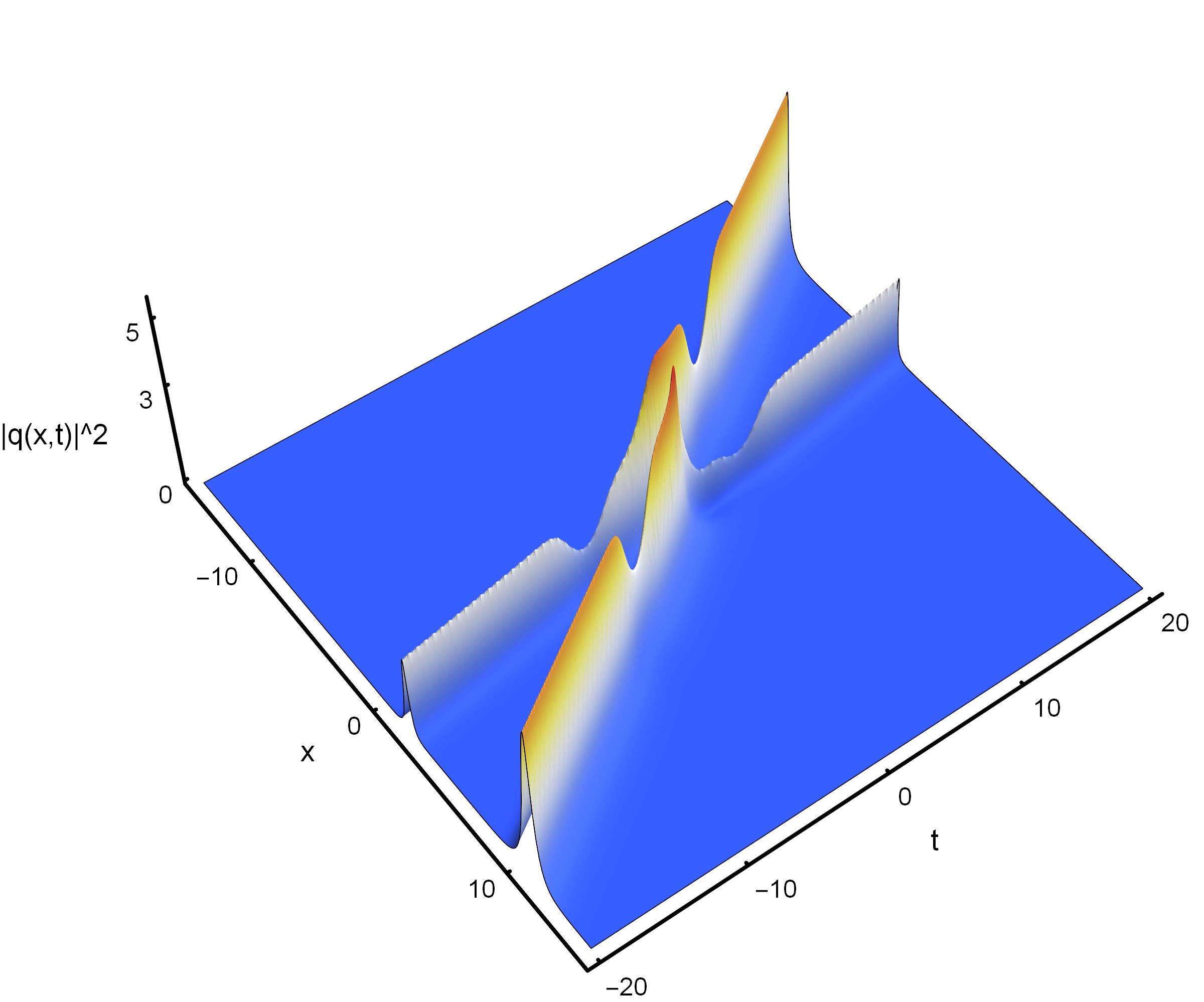}%
		\label{fig:5d}%
	}
		\caption{\scriptsize{3D plot of (a) a 1SS tunneling through double dispersion barrier $D(t)=0.5+e^{-0.5(t-6)^2}+e^{-0.5(t+6)^2}$, $R(t)=1$, with $p_1=1+i$, $\alpha_1=5$, and (b) shows the tunneling of 1SS through double nonlinear barrier $R(t)=0.5+e^{-0.5(t-6)^2}+ e^{-0.5(t+6)^2}$, $D(t)=1$, $p_1=1+i$, $\alpha_1=5$. In (c) and (d) we show the 2SS tunneling through dispersion $D(t)=0.5+0.5e^{-0.5(t-6)^2}+0.5e^{-0.5(t+6)^2}$, $R(t)=1$ and nonlinearity barrier $R(t)=0.5+0.5e^{-0.5(t-6)^2}+0.5 e^{-0.5(t+6)^2}$, $D(t)=1$ respectively with the same parametric values as $p_1=1+i$, $p_2=2+2i$, $\alpha_1=5+0.5i$, $\alpha_2=-5-0.7i$. The loss and gain coefficients for (a) and (b) are $\Gamma(t)=\frac{- e^{-0.5 (t-6)^2} (t-6)- e^{-0.5 (t+6)^2} (t+6)}{2 \left(0.5\, +e^{-0.5 (t-6)^2}+e^{-0.5 (t+6)^2}\right)}$ and $\Gamma(t)=\frac{ e^{-0.5 (t-6)^2} (t-6)+ e^{-0.5 (t+6)^2} (t+6)}{2 \left(0.5\, +e^{-0.5 (t-6)^2}+e^{-0.5 (t+6)^2}\right)}$, respectively. The loss and gain coefficients for (c) and (d) are $\Gamma(t)=\frac{-0.5 e^{-0.5 (t-6)^2} (t-6)-0.5 e^{-0.5 (t+6)^2} (t+6)}{2 \left(0.5\, +0.5 e^{-0.5 (t-6)^2}+0.5e^{-0.5 (t+6)^2}\right)}$ and $\Gamma(t)=\frac{0.5 e^{-0.5 (t-6)^2} (t-6)+0.5 e^{-0.5 (t+6)^2} (t+6)}{2 \left(0.5\, +0.5e^{-0.5 (t-6)^2}+0.5e^{-0.5 (t+6)^2}\right)}$, respectively.}}
	\label{fig:5}
\end{figure}
\indent Now, we  choose the nonlinearity barrier as $R(t)=1+e^{-0.1t^2}$, $D(t)=1$ and $\Gamma(t)=\frac{0.1 te^{-0.1t^2}}{1+ e^{-0.1 t^2}}$ as seen in Figs. \ref{fig:3b}, \ref{fig:3d}. The 2SS propagation through nonlinearity barrier is shown in Figs. \ref{fig:4b}, \ref{fig:4d}. The soliton experiences a constant nonlinearity of strength $1$ and then the second term is again responsible for the Gaussian barrier. Unlike the dispersion barrier which attains a peak at $t=t_0$, the nonlinearity barrier generates a valley at $t=t_0$. The valley arises due to the gain parameter $\Gamma(t)$ which is zero at the location of the valley. Note that the sign of $\Gamma(t)$ is negative for dispersion barrier and positive in case of nonlinearity barrier resulting in peak and valley respectively.\\
\indent Now, we extend the analysis to investigate the propagation of soliton through double Gaussian peaks or valleys. For this purpose, we consider
$D(t)=0.5+e^{-0.5(t-6)^2}+e^{-0.5(t+6)^2}$, $R(t)=1$ that allows the 1SS to travel through double Gaussian peaks at $t=t_0=\pm 6$ as seen in Fig. \ref{fig:5a}.   
The loss parameter in this case is $\Gamma(t)=\frac{- e^{-0.5 (t-6)^2} (t-6)- e^{-0.5 (t+6)^2} (t+6)}{2 \left(0.5\, +e^{-0.5 (t-6)^2}+e^{-0.5 (t+6)^2}\right)}$ and the height of each of the peaks is unity and width is $0.5$ each respectively. The propagation of 2SS through double Gaussian peaks is also obtained when $D(t)=0.5+0.5e^{-0.5(t-6)^2}+0.5 e^{-0.5(t+6)^2}$, $R(t)=1$ as shown in Fig. \ref{fig:5c}. The heights and widths of two peaks are $0.5$ each. In this case the loss parameter is $\Gamma(t)=\frac{-0.5 e^{-0.5 (t-6.)^2} (t-6)-0.5 e^{-0.5 (t+6)^2} (t+6)}{2 \left(0.5\, +0.5e^{-0.5 (t-6.)^2}+0.5e^{-0.5 (t+6)^2}\right)}$. 
The peaks are formed at $t=t_0=\pm6$ and both the 1SS and 2SS experience a constant dispersion of $0.5$. 
%The, dispersion barrier in this case appears as Gaussian double humps at $t=\pm 2$. Suppose, if we consider $D(t)=\Delta_0+\Delta_1 e^-(t+t_0)^2+\Delta_2 e^-(t-t_0)^2$, then $\Delta_0$ is the constant dispersion responsible for constant amplitude of the soliton, $\Delta_1$ and $\Delta_2$ are the barrier heights at $-t_0$ and $t_0$ respectively. For the case, $\Delta_1=\Delta_2$ the heights of the two barrier humps become equal which is evident from the figures \ref{fig:1c} and \ref{fig:1f}. Note that, the amplitude of the envelopes are same in figures \ref{fig:1b}-\ref{fig:1c} and \ref{fig:1e}-\ref{fig:1f} since, barrier heights $\Delta_0$, $\delta_1$ are equal in both the cases and $\Delta_1=\Delta_2$. Only the Gaussian barrier multiplies in the later, as energy accumulates in two regions namely, $t_0=\pm 2$. The soliton tunnels through two barriers in this scenario.  
 We consider the same parametric values of  1SS and 2SS propagating through double Gaussian valleys to investigate the tunneling of 1SS, 2SS through nonlinearity barrier as seen in Figs. \ref{fig:5b}, \ref{fig:5d}. It is evident from the figures that two valleys arise at the same location of the two peaks in the corresponding Figs. \ref{fig:5a}, \ref{fig:5c}.
%%%%%%%%%%%%%%%%%%%%%%%%%%%%%%%%%%%%%%%%%%%%%%%%%%%%%%%%%%%%%%%%%%%%%%%%%%%%%%%%%%%%%%%%%%%%%%%%%%%%%%%%%%%%%%%%%%%%%%%%%%%%%%%%%%%%%%%%%%%%%%%%%%%%%%%%%%%%%%%%%%%%%%%%%%%%%%%%%%%%%%%%
\section{Conclusion}
In this study, we investigated the nonlinear dynamics of vcFLE that controls the evolution of ultrashort optical pulses in inhomogeneous fibre systems. The associated Lax pair is also given which ensures the integrability of the given vcFLE system. The first three conserved quantities are also obtained and modulation instability analysis is performed. By employing  nonstandard Hirota bilinearization method we obtain the bright one soliton, two soliton solution and also provide the scheme to obtain the $N$-bright soliton solution by using auxiliary function. We have also shown that the obtained bright solitons undergo elastic collision with phase shift by systematically performing asymptotic analysis and calculating the transition amplitude and intensity redistribution. We have also obtained the relative separation distance between the two solitons before and after interaction. \\
Subsequently, under various physical scenarios of dispersion and nonlinearity we have also studied the acceleration or retardation of solitons and also the dramatic concept of nonlinear tunneling of 1SS, 2SS through dispersion peaks and nonlinear valleys. The solitons maintain their shape after tunneling through the peak or valleys. Consequently, in this paper we have given a full fledged study of the vcFLE bright soliton dynamics by incorporating most of the physical aspects. We believe that the present study on vcFLE can be helpful for many future studies related to dispersion managed systems, nonlinear tunneling, etc. Additionally, the obtained solutions are also expected to offer qualitative insight into experimentally observed phenomena such as soliton pulsations\cite{si2024deep}, switching mechanism for soliton molecules\cite{si2025polarization}, etc. The presence of variable dispersion, nonlinearity, and gain/loss allows the model to capture key features of realistic laser cavities. Hence, we believe the present results also may serve as a useful analytical reference for interpreting and guiding future experimental studies in ultrafast photonic systems. 
%%%%%%%%%%%%%%%%%%%%%%%%%%%%%%%%%%%%%%%%%%%%%%%%%%%%%%%%%%%%%%%%%%%%%%%%%%%%%%%%%%%%%%%%%%%%%%%%%%%%%%%%%%%%%%%%%%%%%%%%%%%%%%%%%%%%%%%%%%%%%%%%%%%%%%%%%%%%%%%%%%%%%%%%%%%%%%%%%%%%%%%%%%            
\begin{acknowledgements}
S.T. wants to thank DST, Govt. of India for INSPIRE Fellowship (Award No. DST/INSPIRE Fellowship/2020/IF200278).
R.R. would like to acknowledge the financial support in the form of DST-ANRF National Post-doctoral Fellowship (File No. PDF/2023/001115).
M.L. acknowledges DST-ANRF, India for the award of a DST-ANRF National Science Chair (NSC/2020/000029).
\end{acknowledgements}

\section*{Data availability} The authors declare that all data generated or analyzed
during this study are included in this article.

 \section*{Conflict of interest}
 The authors declare that they have no conflict of interest
 concerning the publication of this manuscript.

\bibliographystyle{spphys}       % APS-like style for physics
\bibliography{references}   % name your BibTeX data base

\begin{thebibliography}{10}
\providecommand{\url}[1]{{#1}}
\providecommand{\urlprefix}{URL }
\expandafter\ifx\csname urlstyle\endcsname\relax
  \providecommand{\doi}[1]{DOI \discretionary{}{}{}#1}\else
  \providecommand{\doi}{DOI \discretionary{}{}{}\begingroup
  \urlstyle{rm}\Url}\fi

\bibitem{Hasegawa}
A.~Hasegawa, F.~Tappert, Applied Physics Letters \textbf{23}(3), 142 (1973)

\bibitem{mollenauer1980experimental}
L.F. Mollenauer, R.H. Stolen, J.P. Gordon, Physical Review Letters
  \textbf{45}(13), 1095 (1980)

\bibitem{wang2025launching}
D.~Wang, Z.~Liu, H.~Zhao, H.~Qin, G.~Bai, C.~Chen, P.~Shi, Y.~Du, Y.~Zhao,
  W.~Liu, et~al., Science \textbf{389}(6763), 935 (2025)

\bibitem{xu2025multipole}
T.Z. Xu, J.H. Liu, C.Q. Dai, X.Y. Wang, Nonlinear Dynamics \textbf{113}(23),
  32713 (2025)

\bibitem{serkin2000novel}
V.N. Serkin, A.~Hasegawa, Physical Review Letters \textbf{85}(21), 4502 (2000)

\bibitem{SSE}
N.~Sasa, J.~Satsuma, Journal of the Physical Society of Japan \textbf{60}(2),
  409 (1991)

\bibitem{Kaup}
D.J. Kaup, A.C. Newell, Journal of Mathematical Physics \textbf{19}(4), 798
  (1978)

\bibitem{kivshar2003optical}
Y.S. Kivshar, G.P. Agrawal, \emph{Optical solitons: from fibers to photonic
  crystals} (Academic press, 2003)

\bibitem{agrawal2000nonlinear}
G.P. Agrawal, \emph{Applications of Nonlinear Fiber Optics} (Academic Press,
  San Diego, 2001)

\bibitem{hasegawa1995solitons}
A.~Hasegawa, Y.~Kodama, \emph{Solitons in optical communications} (Oxford
  University Press, 1995)

\bibitem{zabusky1965interaction}
N.J. Zabusky, M.D. Kruskal, Physical Review Letters \textbf{15}(6), 240 (1965)

\bibitem{kuznetsov1986soliton}
E.~Kuznetsov, A.~Rubenchik, V.E. Zakharov, Physics Reports \textbf{142}(3), 103
  (1986)

\bibitem{mou2025optical}
D.S. Mou, Z.Z. Si, W.X. Qiu, C.Q. Dai, Optics \& Laser Technology \textbf{181},
  111774 (2025)

\bibitem{serkin2002exactly}
V.N. Serkin, A.~Hasegawa, IEEE Journal of selected topics in Quantum
  Electronics \textbf{8}(3), 418 (2002)

\bibitem{serkin2007nonautonomous}
V.~Serkin, A.~Hasegawa, T.~Belyaeva, Physical Review Letters \textbf{98}(7),
  074102 (2007)

\bibitem{serkin2010solitary}
V.~Serkin, A.~Hasegawa, T.~Belyaeva, Journal of Modern Optics
  \textbf{57}(14-15), 1456 (2010)

\bibitem{kumar2021dispersion}
V.~Kumar, A.~Patel, Optik \textbf{242}, 166648 (2021)

\bibitem{Fokas}
A.~Fokas, Physica D: Nonlinear Phenomena \textbf{87}(1-4), 145 (1995)

\bibitem{Lenells}
J.~Lenells, Studies in Applied Mathematics \textbf{123}(2), 215 (2009)

\bibitem{FL}
J.~Lenells, A.~Fokas, Nonlinearity \textbf{22}(1), 11 (2008)

\bibitem{lashkin2021perturbation}
V.~Lashkin, Physical Review E \textbf{103}(4), 042203 (2021)

\bibitem{he2011self}
J.d. He, J.f. Zhang, Journal of Physics A: Mathematical and Theoretical
  \textbf{44}(20), 205203 (2011)

\bibitem{chen2014peregrine}
S.~Chen, L.Y. Song, Physics Letters A \textbf{378}(18-19), 1228 (2014)

\bibitem{lenells2010dressing}
J.~Lenells, Journal of Nonlinear Science \textbf{20}(6), 709 (2010)

\bibitem{Matsuno1}
Y.~Matsuno, Journal of Physics A: Mathematical and Theoretical \textbf{45}(23),
  235202 (2012)

\bibitem{Matsuno2}
Y.~Matsuno, Journal of Physics A: Mathematical and Theoretical \textbf{45}(47),
  475202 (2012)

\bibitem{talukdar2023multi}
S.~Talukdar, R.~Dutta, G.K. Saharia, S.~Nandy, Journal of Optics
  \textbf{53}(5), 4150 (2023)

\bibitem{dutta2023fokas}
R.~Dutta, S.~Talukdar, G.K. Saharia, S.~Nandy, Optical and Quantum Electronics
  \textbf{55}(13), 1183 (2023)

\bibitem{lu2012novel}
X.~L{\"u}, B.~Tian, Europhysics Letters \textbf{97}(1), 10005 (2012)

\bibitem{liu2022fokas}
S.z. Liu, J.~Wang, D.j. Zhang, Studies in Applied Mathematics \textbf{148}(2),
  651 (2022)

\bibitem{kundu2010two}
A.~Kundu, Journal of Mathematical Physics \textbf{51}(2) (2010)

\bibitem{lu2013nonautonomous}
X.~L{\"u}, M.~Peng, Chaos: An Interdisciplinary Journal of Nonlinear Science
  \textbf{23}(1) (2013)

\bibitem{silem2023nonautonomous}
A.~Silem, J.~Lin, N.~Akhtar, Journal of Physics A: Mathematical and Theoretical
  \textbf{56}(36), 365201 (2023)

\bibitem{wang2017higher}
Z.Q. Wang, X.~Wang, L.~Wang, W.R. Sun, F.H. Qi, Superlattices and
  Microstructures \textbf{102}, 189 (2017)

\bibitem{serkin2001nonlinear}
V.~Serkin, V.~Chapela, J.~Percino, T.~Belyaeva, Optics Communications
  \textbf{192}(3-6), 237 (2001)

\bibitem{serkin2013geiger}
V.~Serkin, A.~Hasegawa, T.~Belyaeva, Journal of Modern Optics \textbf{60}(2),
  116 (2013)

\bibitem{dutta2025soliton}
R.~Dutta, G.K. Saharia, S.~Talukdar, S.~Nandy, Optical and Quantum Electronics
  \textbf{57}(2), 163 (2025)

\bibitem{Ghosh1999}
S.~Ghosh, A.~Kundu, S.~Nandy, Journal of Mathematical Physics \textbf{40}(4),
  1993 (1999)

\bibitem{mlbook}
M.~Lakshmanan, S.~Rajasekar, \emph{Nonlinear Dynamics, Integrability, Chaos,
  and Patterns} (Springer-Verlag Heidelberg, 2003)

\bibitem{Hirota_2004}
R.~Hirota, \emph{The Direct Method in Soliton Theory}.
\newblock Cambridge Tracts in Mathematics (Cambridge University Press, 2004)

\bibitem{yang2006cascade}
G.~Yang, R.~Hao, L.~Li, Z.~Li, G.~Zhou, Optics Communications \textbf{260}(1),
  282 (2006)

\bibitem{zhong2010soliton}
W.P. Zhong, M.R. Beli{\'c}, Physical Review E—Statistical, Nonlinear, and
  Soft Matter Physics \textbf{81}(5), 056604 (2010)

\bibitem{newell1978nonlinear}
A.C. Newell, Journal of Mathematical Physics \textbf{19}(5), 1126 (1978)

\bibitem{dai2012controllable}
C.Q. Dai, G.Q. Zhou, J.F. Zhang, Physical Review E—Statistical, Nonlinear,
  and Soft Matter Physics \textbf{85}(1), 016603 (2012)

\bibitem{dai2010nonlinear}
C.~Dai, Y.~Wang, J.~Zhang, Optics Express \textbf{18}(16), 17548 (2010)

\bibitem{mani2013dispersion}
M.~Mani~Rajan, J.~Hakkim, A.~Mahalingam, A.~Uthayakumar, The European Physical
  Journal D \textbf{67}(7), 150 (2013)

\bibitem{musammil2017ultrashort}
N.~Musammil, K.~Porsezian, K.~Nithyanandan, P.~Subha, P.T. Dinda, Optical Fiber
  Technology \textbf{37}, 11 (2017)

\bibitem{nandy2021dark}
S.~Nandy, A.~Barthakur, Chaos, Solitons \& Fractals \textbf{143}, 110560 (2021)

\bibitem{si2024deep}
Z.Z. Si, D.L. Wang, B.W. Zhu, Z.T. Ju, X.P. Wang, W.~Liu, B.A. Malomed, Y.Y.
  Wang, C.Q. Dai, Laser \& Photonics Reviews \textbf{18}(12), 2400097 (2024)

\bibitem{si2025polarization}
Z.Z. Si, Z.T. Ju, L.F. Ren, X.P. Wang, B.A. Malomed, C.Q. Dai, Laser \&
  Photonics Reviews \textbf{19}(2), 2401019 (2025)

\end{thebibliography}

%\begin{thebibliography}{}
% etc
%\end{thebibliography}

\end{document}